\documentclass[%
 aip,jcp,
 amsmath,amssymb,
preprint,%
]{revtex4-1}
\usepackage{graphicx}
\usepackage{dcolumn}
\usepackage{bm}
\usepackage{multirow}
\usepackage[utf8]{inputenc}
\usepackage[T1]{fontenc}
\usepackage{mathptmx}
\usepackage{etoolbox}

\usepackage{hyperref}
\hypersetup{
  colorlinks   = true,    
  urlcolor     = blue,    
  linkcolor    = blue,    
  citecolor    = blue      
}
\usepackage{txfonts}
\usepackage{float}
\usepackage{url}
\usepackage{threeparttable}
\usepackage{booktabs}
\usepackage{enumerate}
\usepackage{longtable}
\newcolumntype{C}{>{$}c<{$}}

\usepackage[version=4]{mhchem}
\usepackage{nicefrac}

\makeatletter
\def\@email#1#2{%
 \endgroup
 \patchcmd{\titleblock@produce}
  {\frontmatter@RRAPformat}
  {\frontmatter@RRAPformat{\produce@RRAP{*#1\href{mailto:#2}{#2}}}\frontmatter@RRAPformat}
  {}{}
}%
\makeatother

\usepackage{xr}
\makeatletter
\newcommand*{\addFileDependency}[1]{
  \typeout{(#1)}
  \@addtofilelist{#1}
  \IfFileExists{#1}{}{\typeout{No file #1.}}
}
\makeatother

\newcommand*{\myexternaldocument}[2][]{%
    \externaldocument[#1]{#2}
    \addFileDependency{#2.tex}
    \addFileDependency{#2.aux}
    \addFileDependency{#2.bbl}
    \addFileDependency{#2Notes.bib}
}
\myexternaldocument[B-]{supplemental}

\begin{document}

\title{A Theory of Pitch for the Hydrodynamic Properties of Molecules, Helices, and Achiral Swimmers at Low Reynolds Number}
\author{Anderson D. S. Duraes}
\affiliation{251 Nieuwland Science Hall, \\
  Department of Chemistry \& Biochemistry, \\
  University of Notre Dame, Notre Dame, Indiana 46556, USA}
\author{J. Daniel Gezelter}
\email{gezelter@nd.edu}
\affiliation{251 Nieuwland Science Hall, \\
  Department of Chemistry \& Biochemistry, \\
  University of Notre Dame, Notre Dame, Indiana 46556, USA}

\date{\today}

\begin{abstract}
\vspace{0.7in}
\begin{center} 
\bfseries \large{Abstract}
\end{center}
\vspace{-0.05in}
We present a theory for pitch, a matrix property which is linked to the coupling of rotational and translational motion of rigid bodies at low Reynolds number. The pitch matrix is a geometric property of objects in contact with a surrounding fluid, and it can be decomposed into three \textit{principal axes of pitch} and their associated \textit{moments of pitch}. The moments of pitch predict the translational motion in a direction parallel to each pitch axis when the object is rotated around that axis, and can be used to explain translational drift, particularly for rotating helices. We also provide a symmetrized boundary element model for blocks of the resistance tensor, allowing calculation of the pitch matrix for arbitrary rigid bodies. We analyze a range of chiral objects, including chiral molecules and helices. Chiral objects with a $C_n$ symmetry axis with $n > 2$ show additional symmetries in their pitch matrices. We also show that some achiral objects have non-vanishing pitch matrices, and use this result to explain recent observations of achiral microswimmers. We also discuss the small, but non-zero pitch of Lord Kelvin's isotropic helicoid.
\end{abstract}

\maketitle

\section{Introduction}


Screws are simple machines which are universally used to join parts together and to provide secure enclosures for containers. We often draw a distinction between mechanical screws, which move through solid media, and screws which move through a fluid medium, such as self-propelled swimmers. In all cases, the function and efficiency of the screw is associated with the translation-rotation coupling in the medium, which converts rotation around an axis to linear motion along that axis. The translation-rotation coupling is quantified by the screw's pitch, which is the distance the screw translates upon completing one revolution. This is an intuitive approach for screws that involve contact between two solid surfaces.\cite{Ball1876} For example, a nut advancing through consecutive threads of a screw travels exactly the distance between threads in a single $2\pi$-radian revolution.

For swimmers and other hydrodynamic screws, where physical threads are not explicit, pitch may be an empirically-measured quantity that is challenging to obtain.\cite{Van2021} When the screw operates in a fluid, like a propeller operating in seawater, the translation-rotation coupling decreases because of slip along the screw's surface.\cite{Baranova1978}
Schamel \textit{et al}.\cite{Fischer2013} and Patil \textit{et al}.\cite{Patil2021} observed this fact when studying helical systems moving in liquid media: the translational motion of the helices in a $2\pi$-revolution was less than the standard pitch definition for a helix, which is the distance between consecutive helical turns.

Translational motion can be generated either by rotating the screw itself or induced by rotating the medium around the screw. Howard \textit{et al.} first observed the translational motion of macroscopic chiral objects induced by the vorticity of a fluid.\cite{Howard1976} In their experiment, they suspended dextro-tartaric acid crystals on one side of a drum that was filled with Isopar H (isoparaffinic hydrocarbons). Then, they rotated the drum and observed the migration of the crystals due to the vorticity of the fluid. Although tartaric acid crystals do not look like traditional screws, left- and right-handed crystals exhibit opposite signs of the translational-rotational portion of their resistance tensors, which govern frictional forces and torques experienced by a body in a fluid. This behavior can be used to separate bodies of opposing handedness, since they move in opposite directions in response to rotations. In an earlier paper, we explored the link between parts of the resistance tensor and the translational-rotational coupling, and derived a geometric quantity we called ``scalar pitch''.\cite{Duraes2021} The scalar pitch is a rotational invariant for rigid bodies, describing how the object will translate in response to rotations.
 
In this paper, we extend the concept of pitch that was developed in Ref.~\onlinecite{Duraes2021} into a matrix form, and we investigate the physical properties of the characteristic eigenvalues and eigenvectors of this pitch matrix. This results in a method in which chiral (and achiral) objects can be assigned \textit{principal axes of pitch}, and three associated \textit{moments of pitch} for motion around those axes.  In previous work, the method used for computing resistance tensors was aimed at studying pitch in molecules, so spherical beads representing the atoms were the primary hydrodynamic elements.\cite{Duraes2021}  In this paper, we develop a symmetrized boundary element method using triangular surface patches to evaluate resistance tensors for arbitrary shapes.

\section{Formalism}

\subsection{Resistance and Mobility Tensors}

Consider an arbitrarily shaped rigid body moving in a fluid at low Reynolds number. This rigid body will feel a force and torque in response to its velocity and angular velocity in the fluid. For example, a propeller placed in a flowing fluid experiences a torque, while a screw rotating through a quiescent medium experiences a linear force. 

We define a coordinate system whose origin, $O$, is moving with the body. From Brenner's fundamental work on hydrodynamics,\cite{Brenner1964,Garcia1980} the relationship among net force $(\mathbf{f})$, torque $\left(\bm{\tau}\right)$, velocity $\left(\mathbf{v}\right)$ and angular velocity $\left(\bm{\omega}\right)$ at $O$ is:
\begin{equation}
\left[\begin{array}{c}
\mathbf{f}\\
\bm{\tau}
\end{array}\right]= - \, \Xi \, \left[\begin{array}{c}
\mathbf{v}\\
\bm{\omega}
\end{array}\right]\,,\; \textrm{where}\;\;
\Xi = \left[\begin{array}{cc}
\Xi^{\mathrm{tt}} & \Xi^{\mathrm{rt}}_O\\
\Xi^{\mathrm{tr}}_O & \Xi^{\mathrm{rr}}_O
\end{array}\right]\;.
\label{eq:resistance_tensor}
\end{equation}
$\Xi$ is a $6 \times 6$ hydrodynamic resistance tensor that provides details on how the body couples to the surrounding medium. The four blocks of $\Xi$ represent the translational (tt), rotational (rr), translation-rotation (tr) and rotation-translation (rt) coupling of the body to the medium. The (rt) coupling is the matrix transpose of the (tr) coupling, $\displaystyle{\Xi^{\mathrm{rt}}_O = \left(\Xi^{\mathrm{tr}}_O\right)^T}$, and the subscript $O$ indicates the quantities which depend on the location of the reference point $O$. 

The inverse of the resistance tensor is known as the mobility tensor,\cite{Brenner1967}
\begin{equation}
\label{eq:mobility_tensor}
\bm{\mu}=\left[\begin{array}{cc}
{\mu}_{O}^\mathrm{tt} & {\mu}_{O}^\mathrm{rt}\\
{\mu}_{O}^\mathrm{tr} & {\mu}^\mathrm{rr}
\end{array}\right]= \Xi^{-1}\,.
\end{equation}
$\bm{\mu}$ also comprises four blocks that are analogous to the blocks of the hydrodynamic resistance tensor $\Xi$ defined in Eq.~\eqref{eq:resistance_tensor}. Multiplying the mobility tensor by $k_B\,T$ yields the diffusion tensor, which is a generalization~\cite{Garcia1980,Brenner1967} of Einstein's relation,\cite{Einstein1956} connecting the resistance and diffusion tensors.


\subsection{The Pitch Matrix}

Imagine a simple screw advancing through a material. Because it is coupled to the surrounding medium, if the screw rotates by an angle $\phi$ around its long axis ($z$), it moves linearly along the same rotation axis,
\begin{equation}
\Delta z =  \frac{P}{2 \pi} \,\Delta \phi\,.
\label{eq:pitch_pos}
\end{equation}
This defines the pitch $P$ of the screw in terms of a full $2 \pi$ rotation in $\phi$. 

For continuous rotation in a medium, we can similarly define the pitch in terms of linear and angular velocities of the screw as it rotates around a single axis ($z$),
\begin{equation}
v_z = \frac{P}{2 \pi} \,\omega_z
\label{eq:pitch_vel}
\end{equation}

More generally, the screw may be moving with a (space-fixed) angular velocity vector ($\bm{\omega}$) and the resultant motion may also be a linear velocity vector ($\mathbf{v}$).  In this case, the relationship between $\bm{\omega}$ and $\mathbf{v}$ is mediated by a $3 \times 3$ pitch matrix, 
\begin{equation}
\mathbf{v}=\frac{\mathsf{P}}{2\pi}\,\bm{\omega}
\label{eq:pitch_matrix_generalization}
\end{equation}

Modeling a rigid body as a power screw (which is driven solely by an imposed torque) implies a net force $\mathbf{f}=0$ in Eq. \eqref{eq:resistance_tensor}.\cite{Bhandari2010,Shigley2011} We can then equate the drag force from translational motion to the rotational contribution of the force on the object,
\begin{equation}
    \Xi^{\mathrm{tt}}~\mathbf{v} = -\, \Xi^{\mathrm{rt}}_O~\bm{\omega}
\end{equation}
We can then use the definition of the pitch matrix in Eq.~\eqref{eq:pitch_matrix_generalization}, and obtain an expression in terms of two blocks of the resistance tensor,
\begin{equation}
\Xi^{\mathrm{tt}}~\mathbf{v} = \Xi^{\mathrm{tt}} ~ \frac{\mathsf{P}_O}{2\pi}~ \bm{\omega} = -\, \Xi^{\mathrm{rt}}_O~ \bm{\omega}
\end{equation}
which implies 
\begin{equation}
\frac{\mathsf{P}_O}{2\pi} = - \, \left(\Xi^{\mathrm{tt}}\right)^{-1} \, \Xi^{\mathrm{rt}}_O
\label{eq:pitch_matrix_resistance_tensor}
\end{equation}
where $\displaystyle{\frac{\mathsf{P}_O} {2 \pi}}$ can be seen as a quantity that also depends on the point $O$.

It is also possible to write an equivalent expression for the pitch matrix in Eq.~\eqref{eq:pitch_matrix_resistance_tensor} using two blocks of the mobility tensor. From Eq.~\eqref{eq:mobility_tensor}, the relation between the resistance and mobility tensors can be rewritten as
\begin{equation}
\label{eq:diffusion_tensor_rewritten}
\Xi\,\bm{\mu}=\left[\begin{array}{cc}
\mathbf{I} & \mathbf{0}\\
\mathbf{0} & \mathbf{I}
\end{array}\right]\,,
\end{equation}
where $\mathbf{I}$ is the $3 \times 3$ identity matrix and $\mathbf{0}$ is the $3 \times 3$ null matrix. In terms of the mobility tensor blocks, the pitch matrix is:
\begin{equation}
\frac{\mathsf{P}_O}{2\pi} = {\mu}^{\mathrm{rt}}_O\,\left({\mu}^{\mathrm{rr}}\right)^{-1} 
\label{eq:pitch_matrix_mobility_tensor}
\end{equation}
Note that Eqs. \eqref{eq:pitch_matrix_resistance_tensor} and \eqref{eq:pitch_matrix_mobility_tensor} are equivalent forms. 

We note that when a rigid body is settling under an external force, so that translational motion generates all rotation, we must invoke a different process than a power screw, as the force on the body is no longer zero. In this case, we have no external torque $(\bm{\tau}=0)$ and $\bm{\omega}=\mathsf{L}\,\mathbf{v}$, where $\mathsf{L}$ is a $3\times3$ matrix that mediates the generation of angular velocity from linear velocity.  Using the resistance or mobility tensors, 
\begin{equation}
    \mathsf{L} = -\, \left(\Xi^{\mathrm{rr}}_O\right)^{-1} \Xi^{\mathrm{tr}}_O = {\mu}_{O}^\mathrm{tr}\, \left({\mu}_{O}^\mathrm{tt}\right)^{-1}~,
\end{equation}
which can be related to the pitch matrix defined in Eqs.~\eqref{eq:pitch_matrix_resistance_tensor} or \eqref{eq:pitch_matrix_mobility_tensor},
\begin{equation}
\frac{\mathsf{P}_O} {2 \pi} = \left(\Xi^{\mathrm{tt}}\right)^{-1}\, \mathsf{L}^{T} \, \left(\Xi^{\mathrm{rr}}_O\right) = {\mu}_{O}^\mathrm{tt} \; \mathsf{L}^{T} \, \left({\mu}^\mathrm{rr}\right)^{-1}.
\end{equation}  
Ekiel-Je{\.{z}}ewska and Wajnryb\cite{Ekiel-Jezewska2009} studied this process  using a three-sphere (trumbbell) model settling under gravity in a viscous fluid and concluded that the trumbbell rotates as it settles. In Sec.~\ref{subsec:isotropic_helicoid}, we consider a similar example using an isotropic helicoid falling through a fluid, where its linear velocity induces a small angular velocity.





\subsection{Center of Pitch \label{subsec:center_of_pitch}}

The translational (tt) and rotational (rr) blocks of the resistance (Eq.~\eqref{eq:resistance_tensor}) and mobility (Eq.~\eqref{eq:mobility_tensor}) tensors are symmetric matrices for any point $O$. However, the blocks which couple translation and rotation are only symmetric at the center of resistance (CR) for the resistance tensor,\cite{Brenner1964,Garcia1980,Duraes2021} and at the center of diffusion (CD) for the mobility tensor.\cite{Garcia1980,Duraes2021} Therefore, the pitch matrix is not generally a symmetric matrix.  However, at one special point, which we call the \textit{center of pitch} ($\mathbf{p}$), the pitch matrix does become symmetric.

For the resistance tensor, the translation-rotation couplings at a point $M$ (separate from the origin) will include a portion of the translational block along the line connecting the origin $O$ to $M$, while the translation-rotation couplings for the mobility tensor will include a portion of the rotational block.\cite{Brenner1964,Garcia1980,Duraes2021} We can express the new couplings,
\begin{eqnarray}
\label{eq:resistance_tr_coupling_transform}
 \Xi_M^\mathrm{tr}  & = & \Xi_O^\mathrm{tr} \, - \,\mathsf{U}_{M} \,\Xi^\mathrm{tt} \\
 \label{eq:diffusion_tr_coupling_transform}
 {\mu}_{M}^\mathrm{tr}  & = & {\mu}_{O}^\mathrm{tr} \, + \, {\mu}^\mathrm{rr} \, \mathsf{U}_{M}
\end{eqnarray}
where $\mathsf{U}_{M}$ is a skew-symmetric matrix whose elements are set by the vector from point $O$ to point $M$,
\begin{equation}
\mathsf{U}_{M}  = \left[ \begin{array}{ccc}
  \phantom{-}\,0 & -\,z_{OM} & \phantom{-}\,y_{OM} \\
\phantom{-}\,z_{OM} &  \phantom{-}\,0   & -\,x_{OM} \\
-\,y_{OM} & \phantom{-}\,x_{OM} & \phantom{-}\,0
\end{array} \right] \,.
\label{eq:skew_symmetric_matrix}
\end{equation}
Note that $\left(\mathsf{U}_{M}\right)^T = -\,\mathsf{U}_{M}$.

Left-multiplying the transpose of Eq.~\eqref{eq:resistance_tr_coupling_transform} by $\left(\Xi^{\mathrm{tt}}\right)^{-1}$, or right-multiplying the transpose of Eq.~\eqref{eq:diffusion_tr_coupling_transform} by $\left({\mu}^{\mathrm{rr}}\right)^{-1}$, we can see how to transform a pitch matrix computed at a point $O$ to another point $M$,
\begin{equation}
\label{eq:pitch_matrix_transform}
\frac{\mathsf{P}_{M}}{2\pi} = \frac{\mathsf{P}_{O}}{2\pi} \, -\, \mathsf{U}_{M} 
\end{equation}

To find the center of pitch, or the point $\mathbf{p}$ where the pitch matrix is symmetric, we set the right side of Eq.~\eqref{eq:pitch_matrix_transform} equal to its transpose and we find the coordinates of the vector $\mathbf{r}_{\mathbf{p}}$ connecting the center $O$ to $\mathbf{p}$:
\begin{align}
\mathbf{r}_{\mathbf{p}} & =\frac{1}{4 \pi}\left[\begin{array}{c}
\left(\mathsf{P}_O \right)_{zy}-\,\left(\mathsf{P}_O \right)_{yz}\\
\left(\mathsf{P}_O \right)_{xz}-\,\left(\mathsf{P}_O \right)_{zx}\\
\left(\mathsf{P}_O \right)_{yx}-\,\left(\mathsf{P}_O \right)_{xy}
\end{array}\right]\nonumber 
\end{align}
where the subscripts indicate the entries of the $\displaystyle{\mathsf{P}_{O}}$ matrix. The  supplementary material (Sec.~\ref{B-sec:proof_uniqueness_center_of_pitch})  provides an additional proof that the center of pitch is unique to each body.

The symmetric pitch matrix can be found without knowing the center of pitch,
\begin{equation}
\mathsf{P}_\mathbf{p} = \frac{1}{2}\,\left[\mathsf{P}_{O} \, + \, \mathsf{P}_{O}^T\right] .
\end{equation}
However, this relation does not provide the location of the center of pitch.

\subsection{Pitch Axes, Moments of Pitch, and The Pitch Coefficient}
From the original definition of pitch (Eq.~\eqref{eq:pitch_matrix_generalization}), we can diagonalize\cite{Riley_Hobson_Bence2006} the symmetric pitch matrix and write
\begin{equation}
\mathsf{A}^{-1} \left( \frac{\mathsf{P}_\mathbf{p}}{2\pi} \right) \mathsf{A} = \left[\begin{array}{ccc} \lambda_1 & 0 & 0 \\
0 & \lambda_2 & 0 \\
0 & 0 & \lambda_3 \end{array} \right]
    \end{equation}
where the three eigenvalues ($\lambda_i$) are \textit{moments of pitch} --- each associated with a \textit{pitch axis}, $\mathbf{\alpha}_i$ --- which is one of the column vectors making up $\mathsf{A} = \left[ \mathbf{\alpha}_1, \mathbf{\alpha}_2, \mathbf{\alpha}_3 \right]$.
This decomposition into principal axes and moments of pitch is a direct analogy to the decomposition of a moment of inertia tensor into principal axes and moments of inertia. For the pitch matrix, however, moments of pitch may be negative if rotating the body counterclockwise around axis $\mathbf{\alpha}_i$ results in translation along the negative $\mathbf{\alpha}_i$ direction.  

\begin{figure}[b]
\includegraphics[width=0.7\linewidth]{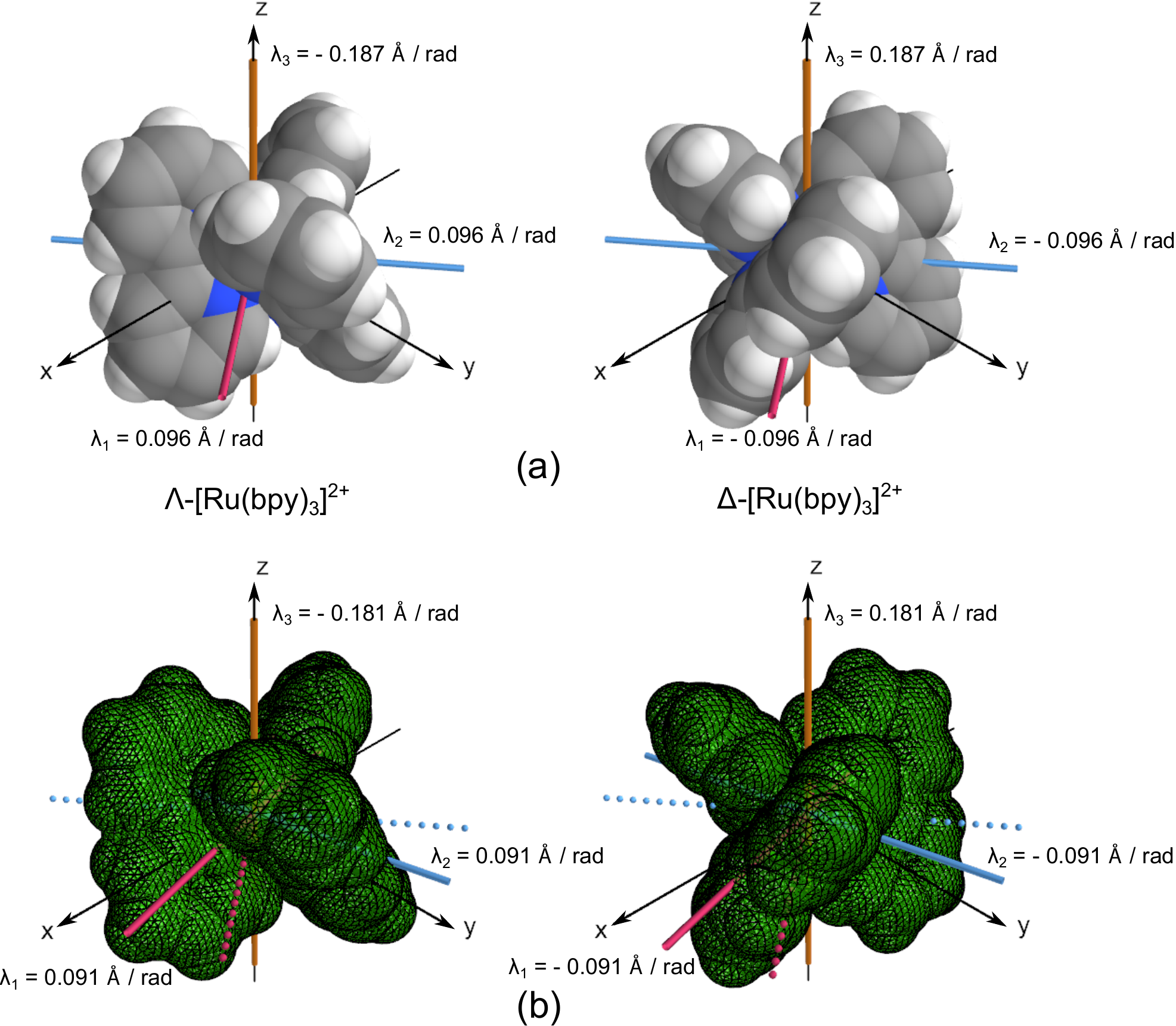}
\caption{The pitch axes and the moments of pitch for the $\Lambda$ and $\Delta$ enantiomers of the \ce{[Ru(bpy)_3]^{2+}} ion. In the upper panel~(a), spheres with radii from the OPLS-AA force field\cite{Jorgensen1996} were used to compute the resistance tensor, while in the lower panel~(b), a triangulated surface mesh was utilized. The two enantiomers have opposing signs for all three of their moments of pitch. Note that two moments of pitch are degenerate, indicating that a linear combination of the corresponding pitch axes will also form a basis along the $xy$-plane (the pink and blue lines or the pair of dotted lines in the lower panel, for example). The pitch axis along the $z$-axis is aligned with the $C_3$ axis of the molecule. The scalar pitch coefficient for this molecule is $|P|/2\pi = 0.133~\AA{} / \mathrm{rad}$.\label{fig:molecules}}
\end{figure}

It is useful to define a rotational invariant which will provide information on the average translational motion exhibited when the body has a random orientation in the fluid. We can define a \textit{scalar pitch coefficient} which is the simplest rotational invariant of the pitch matrix, providing equal contributions from rotation around all three axes of pitch,
\begin{equation} \label{eq:pitch_coefficient}
    \frac{|P|}{2 \pi} = \sqrt{ \frac{1}{3} \sum_i \lambda_i^2}
\end{equation}
The derivation of the \textit{scalar pitch coefficient} is available  in Sec.~\ref{B-sec:pitch_coefficient_generalization} of the supplementary material.  Note that this is functionally equivalent to a pitch coefficient that was demonstrated in our previous paper,\cite{Duraes2021} where the eigenvalues of the (tt) and (tr) blocks were used separately to compute $|P| / 2 \pi$.

Figure \ref{fig:molecules} displays the principal axes of pitch and their associated moments for the $\Lambda$ and $\Delta$ enantiomers of the \ce{[Ru(bpy)_3]^{2+}} ion. If there are degeneracies in the moments of pitch, a linear combination of the corresponding pitch axes will also form a basis for understanding the translation-rotation coupling of the object.  

\subsection{Pitch Properties of Enantiomers and Achiral Objects\label{subsec:pitch_properties}}

Consider a left- and right-handed pair of enantiomers, whose structures are related by a reflection through the origin. For enantiomers, the (tt) and (rr) blocks of the resistance and mobility tensors will be identical, while the (tr) and (rt) blocks flip sign.\cite{Duraes2021} Using this mirror image property, we can deduce the following property of the pitch matrix for the two enantiomers:
\begin{equation}
\left(\frac{\mathsf{P}_O}{2\pi}\right)^{\mathrm{left}} ~=~ - \, \left(\frac{\mathsf{P}_O}{2\pi}\right)^{\mathrm{right}}
\label{eq:pitch_matrix_mirror_image_property}
\end{equation}


At the center of pitch, Eq.~\eqref{eq:pitch_matrix_mirror_image_property} implies that the moments of pitch (eigenvalues) of the pitch matrix for the left- and right-handed objects have the same magnitude, but flip signs. The pitch axes (eigenvectors), however, are identical for both objects.

\begin{figure}[b]
    \centering
    \includegraphics[width=0.75\linewidth]{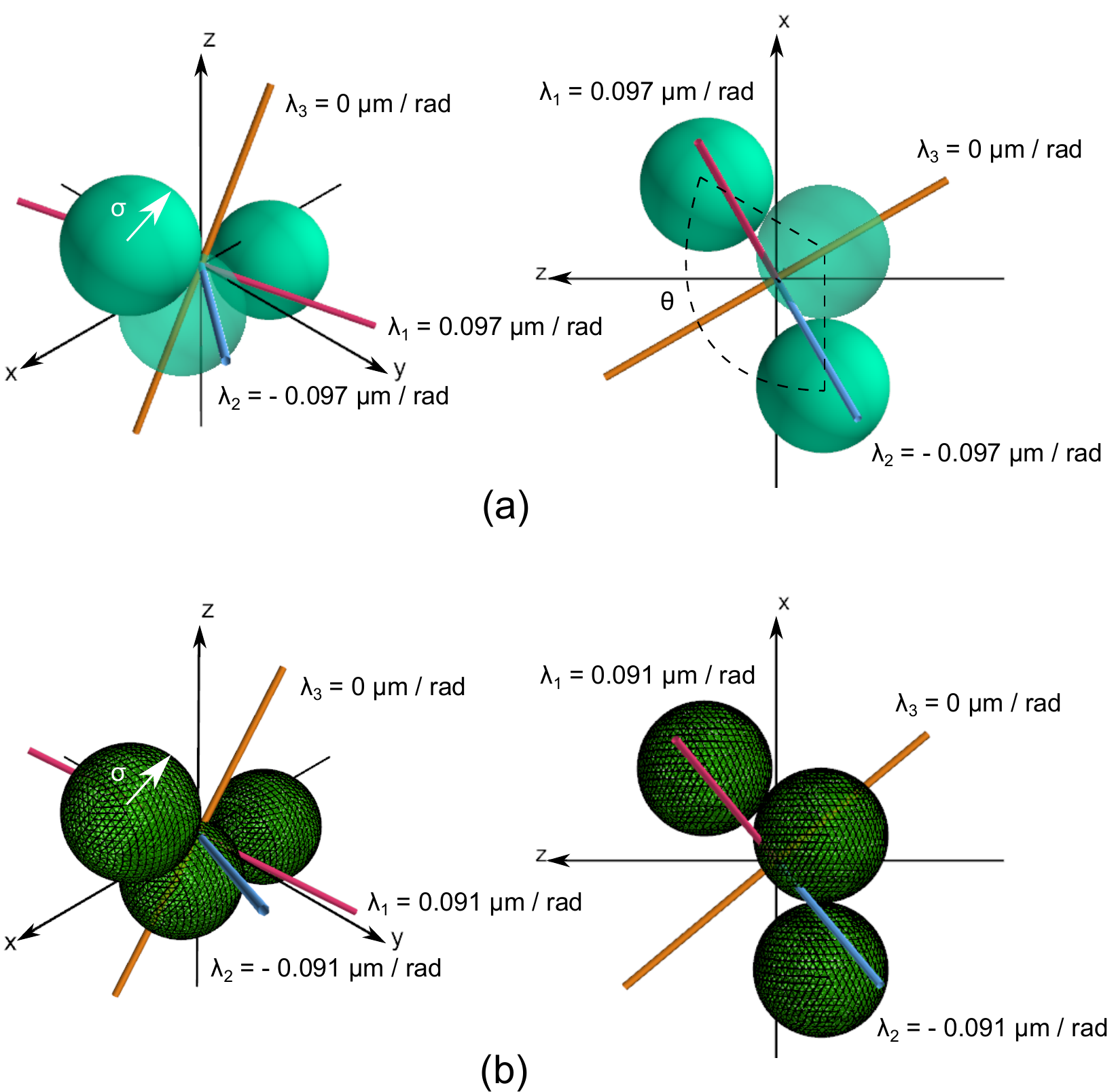}
    \caption{Achiral swimmers, like the three bead arrangement shown here, can exhibit non-zero pitch matrices and two moments of pitch with opposite signs. This means that rotation of these swimmers can result in translation through a fluid. In this collection of particles, each of the three beads has a radius of $\sigma=2.2~\mu \mathrm{m}$ and an angle $\theta = 120^{\circ}$. This corresponds to class~\protect\hyperlink{itm:case_ii_lambda1_n_lambda2}{2} in the classification of achiral objects. In the upper panel~(a), spheres were used to compute the resistance tensor, while in the lower panel~(b), a triangulated surface mesh was utilized. \label{fig:achiral}}
\end{figure}

For an achiral object, which is identical with its own mirror image, the characteristic eigenvalues of the left and right pitch matrices must be the same, and there are two ways for this to happen:

\begin{itemize}
\item[1)] $\lambda_1 = \lambda_2 = \lambda_3 = 0$ \label{itm:case_i_all_lambdas_zero} \\
In this case, the achiral object has a pitch matrix which is the null matrix, and thus does not exhibit displacement due to rotation. This situation occurs in objects with a high degree of internal symmetry, \textit{e.g.}, spheres and ellipsoids.

\item[2)] $\lambda_1 = 0$ and $\lambda_2 = -\lambda_3 \neq 0$ \label{itm:case_ii_lambda1_n_lambda2} \hypertarget{itm:case_ii_lambda1_n_lambda2}{}\\
In this case, the achiral object has a pitch matrix which is non-zero. These objects can exhibit displacement due to rotation, and this property helps explain the recent observation of achiral microswimmers which can be propelled through a fluid via rotation\cite{Fu2014} (see Fig. \ref{fig:achiral}).
\end{itemize}

\noindent For chiral objects, there are two additional cases to consider:
\begin{itemize}
    \item[3)] $\lambda_1 = \lambda_2 \neq \lambda_3$ (two degenerate eigenvalues). \\
    This class of objects is a chiral body in the $C_n$ or $D_n$ point groups with $n \ge 3$.
    \item[4)] $\lambda_1 \neq \lambda_2 \neq \lambda_3$\\
    This is the general case for most chiral objects without higher symmetry axes.
\end{itemize}

To explore the pitch matrix properties of chiral objects, we can consider the chiral point groups $C_n$ and $D_n$.\cite{Huheey1993} Table~\ref{tab:chiral_point_groups_pitch_matrix} shows the moments of pitch (the eigenvalues of the pitch matrix) for a set of representative molecules that belong to the $C_n$ and $D_n$ point groups.\cite{symotter,NIST_database,Antol2014} To compute these moments of pitch, we first constructed the molecular resistance tensors, representing atoms with spheres with appropriate van der Waals radii. The molecular resistance tensors were computed using the methods described in Ref.~\onlinecite{Duraes2021}. The pitch matrix for each molecule was then constructed and diagonalized to obtain the molecular pitch axes (eigenvectors) and the associated moments of pitch (eigenvalues).  

When the symmetry axis of the point groups has $n \geq 3$, we find that two of the moments of pitch are always degenerate (see table \ref{tab:chiral_point_groups_pitch_matrix}). When this degeneracy occurs, the \textit{non}-degenerate moment of pitch is associated with a pitch axis that points directly along the $C_n$ axis of the molecule. This property is expected from the character tables of the $C_n$ and $D_n$ point groups,\cite{Salthouse1972} since one coordinate (\textit{e.g.}, $z$) forms a basis for a \textit{non}-degenerate irreducible representation, and the other two coordinates (\textit{e.g.}, $x$ and $y$) span a doubly-degenerate representation.


\renewcommand{\arraystretch}{0.8}
\setlength{\tabcolsep}{10pt}
\LTcapwidth=\linewidth
\setlength\LTleft{0pt}            
\setlength\LTright{0pt}           
\begin{longtable*}{|c|cccc|}
\caption{Moments of Pitch for representative molecules from the chiral point groups $C_n$ and $D_n$.\cite{symotter,NIST_database,Antol2014} To compute molecular resistance tensors, atoms were represented with spheres with appropriate van der Waals radii, and resistance tensors were computed using the methods described in Ref.~\onlinecite{Duraes2021}. Degenerate moments of pitch are indicated in bold type. \label{tab:chiral_point_groups_pitch_matrix}} \\

\hline
\multirow[m]{2}{*}{Point Group} &  \multirow[m]{2}{*}{Molecule} & \multicolumn{3}{|c|}{Moments of Pitch ($\times~10^{-4}~\AA{}~\mathrm{rad}^{-1}$)} \\ 
\cline{3-5} &  & \multicolumn{1}{|c}{$\lambda_{1}$} & $\lambda_{2}$ & $\lambda_{3}$ \\ \hline
\endfirsthead

\multicolumn{5}{c}{{\bfseries \tablename\ \thetable{} -- continued from previous page}} \\
\hline
 \multirow[m]{2}{*}{Point Group} &  \multirow[m]{2}{*}{Molecule} & \multicolumn{3}{|c|}{Moments of Pitch ($\times~10^{-4}~\AA{}~\mathrm{rad}^{-1}$)} \\ \cline{3-5}
 &  & \multicolumn{1}{|c}{$\lambda_{1}$} & $\lambda_{2}$ & $\lambda_{3}$ \\ \hline
\endhead
\hline \multicolumn{5}{|r|}{{Continued on next page}} \\ \hline
\endfoot
\hline \hline
\endlastfoot

\multirow[m]{6}{*}{$C_1$}	&	1-bromo-1-chloroethane	&	-229.32	&	11.07	&	223.36	\\
	&	2,3-dihydrofuran	&	-210.45	&	-17.25	&	240.34	\\
	&	bromochlorofluoromethane	&	-156.89	&	22.91	&	134.06	\\
    &	D-alanine	&	-404.01	&	60.99	&	375.02	\\
	&	D-serine	&	-384.40	&	2.37	&	338.14	\\
	&	SOClBr	&	-119.43	&	-14.30	&	133.58	\\ \cline{1-5}
\multirow[m]{7}{*}{$C_2$}	&	1,3-dichloroallene	&	-477.66	&	-25.93	&	583.70	\\
	&	2,3-pentadiene	&	-250.60	&	-182.19	&	405.21	\\
	&	\textit{cis}-\ce{[Co(en)2Cl2]+}	&	-306.88	&	91.10	&	145.42	\\ 
	&	hydrazine	&	-324.38	&	-16.05	&	256.58	\\
	&	hydrogen peroxide	&	-273.39	&	-199.04	&	408.42	\\
	&	\ce{Mo(acac)2O2}	&	-1323.32	&	4.85	&	1266.70		\\
	&	titanium dimer	&	-366.93	&	-212.78	&	600.78		\\ \cline{1-5}
\multirow[m]{4}{*}{$C_3$}	&	tris-aminomethane	&	\textbf{-119.09}	&	\textbf{-119.09}	&	178.09		\\
	&	triethylamine	&	-110.35	&	\textbf{36.66}	&	\textbf{36.66}	\\
	&	triphenylmethane	&	\textbf{-1071.89}	&	\textbf{-1071.89}	&	2056.39	\\
	&	triphenylphosphine	&	\textbf{-857.58}	&	\textbf{-857.58}	&	1645.00	\\ \cline{1-5}
$C_4$	&	tetra-aza copper(II)	&	\textbf{-85.70}	&	\textbf{-85.70}	&	110.43	\\ \cline{1-5}
$C_5$	&	\ce{Fe(Me5}-\ce{Cp)(P5)}	&	-13.70	&	\textbf{11.47}	&	\textbf{11.47}	\\ \cline{1-5} 
$C_6$	&	alpha-cyclodextrin	&	\textbf{-9.79}	&	\textbf{-9.79}	&	34.49	\\ \cline{1-5} 
\multirow[m]{3}{*}{$D_2$} &	biphenyl	&	-1730.16	&	605.84	&	1144.82	\\
	&   \textit{trans}-\ce{[Co(en)2Cl2]+}	&	-153.73	&	79.32	&	90.63	\\ 
	&	twistane	&	-92.72	&	20.74	&	40.50	\\ \cline{1-5}
\multirow[m]{3}{*}{$D_3$}	&	guanidinium cation	&	-39.67	&	\textbf{81.99}	&	\textbf{81.99}	\\
	&	tris(en)cobalt(III) &	\textbf{-176.32}	&	\textbf{-176.32}	&	328.17		\\
	&	tris(oxalato)iron(III)	&	\textbf{-402.24}	&	\textbf{-402.24}	&	953.13		\\ \cline{1-5}
$D_4$	&	tetrathiacyclododecane	&	-218.66	&	\textbf{325.53}	&	\textbf{325.53}	\\ \cline{1-5}
\multirow[m]{2}{*}{$D_5$}	&	twisted ferrocene 	&	-14.26	&	\textbf{10.73}	&	\textbf{10.73}	\\
	&	\ce{YbI2(THF)5}	&	\textbf{-180.88}	&	\textbf{-180.88}	&	356.50	\\ \cline{1-5}
$D_6$	&	bis(benzene)chromium	&	-4.43	&	\textbf{3.51}	&	\textbf{3.51}	\\
\end{longtable*}

\subsection{Hydrodynamic Model: Determining the Resistance Tensor from Triangulated Surfaces\label{subsec:triangulated_surface_method}}

The exterior surface of any rigid body moving through a fluid can be described as a surface mesh comprising small, flat triangular patches. Surface triangulation is a widely-researched topic, and we assume here that the object of interest has been expressed in this form. When viscous forces are dominant, \textit{i.e.}, at low Reynolds number, the
velocity of triangle $i$ ($\mathbf{v}_{i}$) is related to the
unperturbed velocity of the fluid ($\mathbf{u}$) via hydrodynamic interaction
tensors ($\mathsf{B}_{i\!j}$), which connect triangular plate $i$ to the forces experienced by all of the triangular plates comprising the surface of the rigid body,\cite{Allison2001}
\begin{equation} \label{eq:fluid_velocity_triangle_i}
\mathbf{v}_{i}= \mathbf{u} - \sum_{j} \mathsf{B}_{i\!j}\,\mathbf{F}_{j}
\end{equation}
Because the hydrodynamic interaction is reciprocal,\cite{Kim1991} we introduce a symmetrized version of~$\mathsf{B}$ which integrates over both triangular patches to obtain  coupling between triangular elements,
\begin{equation}
\mathsf{B}_{i\!j}=\mathsf{B}_{\!ji}=\frac{1}{2}\left[\frac{1}{A_{\!j}}\int_{S_{\!\!j}} \mathsf{T}\left(\mathbf{x}_{i},\,\mathbf{y}\right)\,d\mathbf{y}+\frac{1}{A_{i}}\int_{S_{\!i}}\mathsf{T}(\mathbf{x}_{j},\,\mathbf{y})\,d\mathbf{y} \right]
\label{eq:hydroTens}
\end{equation}
These are integrals over the surfaces $S_{\!\!j}$ and $S_{\!i}$ of triangles $\!j$ and $i$, respectively, and $\mathbf{x}_{i}$ is the centroid (or barycenter) of triangle $i$.  The area of triangle $j$ can be similarly expressed as a surface integral,
\begin{equation}
A_{\!j}=\int_{S_{\!\!j}}\,d\mathbf{y}~.
\end{equation}
The symmetrized form of $\mathsf{B}$ is essential for maintaining the known properties of the resistance tensor (Sec.~\ref{subsec:center_of_pitch}). 

The Oseen tensor connecting points $\mathbf{a}$ and $\mathbf{b}$,\cite{Allison2001,Kim1991}
\begin{equation}\label{eq:Oseen_tensor_points}
\mathsf{T}\left(\mathbf{a},\,\mathbf{b}\right)=\mathsf{T}\left(\mathbf{b},\,\mathbf{a}\right)=\frac{1}{8\pi\,\eta\,|\mathbf{a}\, - \, \mathbf{b}|}\left[\mathbf{I}\, + \,\frac{\left(\mathbf{a}\, - \,\mathbf{b}\right)\,\otimes\,\left(\mathbf{a}\, - \,\mathbf{b}\right)}{|\mathbf{a}\,-\,\mathbf{b}|^{2}}\right]
\end{equation}
provides the coupling through a surrounding fluid with dynamic viscosity $\eta$. The $\otimes$~symbol indicates the outer (tensor) product of two vectors, in this case, $\left(\mathbf{a}-\mathbf{b}\right)$ with itself.

The surface integrals in Eq.~\eqref{eq:hydroTens} can be calculated numerically using a surface quadrature:
\begin{align}
\frac{1}{A_{\!j}}\int_{S_{\!\!j}} \mathsf{T}(\mathbf{x}_{i},\,\mathbf{y})\,d\mathbf{y} & \approx\frac{1}{A_{\!j}}\left[A_{\!j}\,\sum_{k}w_{k}\,\mathsf{T}(\mathbf{x}_{i},\,\mathbf{y}_{k_j})\right]\nonumber \\
 & \approx\sum_{k}w_{k}\,\mathsf{T}(\mathbf{x}_{i},\,\mathbf{y}_{k_j})\,,
\end{align}
where $w_{k}$ is a weight associated with the quadrature point
$\mathbf{y}_{k_j}$ on triangle $j$, and $\mathbf{x}_{i}$ is the centroid of triangle $i$. Using quadrature points and weights, we can therefore rewrite Eq.~\eqref{eq:hydroTens} as:
\begin{equation}
\mathsf{B}_{i\!j}=\frac{1}{2}\,\sum_{k} w_{k}\left[\mathsf{T}(\mathbf{x}_{i},\,\mathbf{y}_{k_{j}}) \, + \, \mathsf{T}(\mathbf{x}_{j},\,\mathbf{y}_{k_{\,i}})\right]
\end{equation}
If not stated otherwise, we employ a 6-point Gaussian quadrature developed by Cowper\cite{Cowper1973} which exactly integrates polynomials of degree 3 and whose points and weights are available in Quadpy.\cite{Quadpy_github} Because the centroid is not a point in this quadrature, there is no singularity in the self interaction ($\mathsf{B}_{ii}$).  

Using all of the $3\times3$ $\mathsf{B}_{i\!j}$ matrices, we can construct a $3N\times3N$
$\mathsf{B}$ supermatrix, where $N$ stands for the total number of triangular plates, and rewrite Eq.~\eqref{eq:fluid_velocity_triangle_i} as,
\begin{equation} \label{eq:fluid_velocity_triangles_matrix_form}
\mathbf{V} - \mathbf{U} = - \mathsf{B} \,\mathbf{F}\,.
\end{equation}
where $\mathbf{V}$, $\mathbf{U}$, and $\mathbf{F}$ are $3N$-dimensional vectors representing the triangles' velocities, unperturbed fluid velocity, and forces on all of the triangles. The solution of Eq.~\eqref{eq:fluid_velocity_triangles_matrix_form} to find the force requires the inverse, $\mathsf{C}=\mathsf{B}^{-1}$,
\begin{equation}\label{eq:}
\mathbf{F}  = -\, \mathsf{C} \left(\mathbf{V} - \mathbf{U}\right),
\end{equation}
and is equivalent to the translational block of the resistance tensor in Eq.~\eqref{eq:resistance_tensor}, after summing over all the triangles to yield the net translational force on the object,
\begin{align*}
\mathbf{f} &= -\left(\sum_{ij} \mathsf{C}_{i\!j} \right) \mathbf{v} \\
& = -\Xi^{\mathrm{tt}} \,\mathbf{v}
\end{align*}
where $\mathsf{C}_{i\!j}$ are the $3 \times 3$ blocks of the $\mathsf{C}$ matrix, and we have assumed that the assembly of triangles is a rigid body, so all triangles have the same velocity relative to the fluid, $\left(\mathbf{v}_{\!j} - \mathbf{u}\right) =\mathbf{v}$. This allows us to identify the translational resistance tensor,\cite{Garcia1981,Carrasco1999,Garcia1999}
\begin{equation}
\Xi^{\mathrm{tt}} =\sum_{ij} \mathsf{C}_{i\!j}
\end{equation}
From the Brenner relations for the (tr) and (rr) blocks of the resistance tensor in a discrete, matrix form,\cite{Brenner1964,Garcia1980,Duraes2021} we also have
\begin{align}
\begin{split}
\Xi_{O}^{\mathrm{tr}} & =\sum_{ij} \mathsf{U}_{i}\,\mathsf{C}_{i\!j}\,,\\
\Xi_{O}^{\mathrm{rr}} & =-\sum_{ij} \mathsf{U}_{i}\, \mathsf{C}_{i\!j}\, \mathsf{U}_{j} 
\end{split}
\end{align}
where $\mathsf{U}_{i}$ is the skew-symmetric matrix defined in Eq.~\eqref{eq:skew_symmetric_matrix} whose entries $x_{Oi}$, $y_{Oi}$ and $z_{Oi}$ are the components of the vector between the origin $O$ and the centroid of the triangle $i$.

Note that in contrast to bead models,\cite{Torre:1983lr,Carrasco1999,Garcia1999,Duraes2021} a boundary element method does not require a volume correction to the rotational block of the resistance tensor, since the boundary element method computes interactions using hydrodynamic elements that have no volume.

With the blocks of the resistance tensor computed at point $O$, it is possible to reconstruct the blocks at another point $M$. The (tt) block is invariant to choice of origins, the (tr) block follows Eq.~\eqref{eq:resistance_tr_coupling_transform}, and the (rr) block requires coupling to the other blocks of $\Xi$,\cite{Brenner1964,Garcia1980,Duraes2021}
\begin{equation}
\label{eq:resistance_rr_transform}
 \Xi_M^\mathrm{rr} = \Xi_O^\mathrm{rr}  
\, + \, \Xi_O^\mathrm{tr}\, \mathsf{U}_{M} \, - \, \mathsf{U}_{M}\, \left( \Xi_O^\mathrm{tr}
\right)^{T} - \, \mathsf{U}_{M}\, \Xi^\mathrm{tt}\, \mathsf{U}_{M}
\end{equation}

 In Sec.~\ref{B-sec:analytical_expressions} of the supplementary material,  we apply the boundary element method developed here to objects whose blocks of the resistance tensor are known analytically. The boundary element method shows good agreement with the analytical values. 

\section{Results}
Applying the triangulated surface boundary element method described in the previous section, we have computed the principal axes of pitch and moments of pitch for a wide array of objects. These objects include common chiral entities like helices, achiral swimmers, and one object of historical curiosity: Lord Kelvin's Isotropic Helicoid. These objects are swimming at low Reynolds number and, if not stated otherwise, we shifted the center of pitch to the origin of the coordinate system.  Wherever possible, we compare the predictions from the pitch matrix to experimental results for similar objects in similar fluid conditions. For Lord Kelvin's Isotropic Helicoid, we also investigate two feasible experiments to assess its rotation-translation coupling. 

\subsection{Chiral Objects}

\subsubsection{Helices\label{subsec:helices}}
The hydrodynamic properties of helices have been studied widely because of their importance in the motion of living cells. Chwang and Wu~\cite{Chwang1971} and Higdon~\cite{Higdon1979} used a helix connected to a spherical head to model the swimming of microorganisms and to find optimum design parameters for efficient propulsion under low Reynolds numbers. Purcell\cite{Purcell1997} approximated the blocks of the resistance tensor in Eq.~\eqref{eq:resistance_tensor} as scalars, reducing the $6 \times 6$ resistance tensor to a $2 \times 2$ tensor, and explored the relation between these scalar values in the coupling of translational and rotational motions of helical systems. To study the swimming properties of \textit{Escherichia coli} bacteria, Chattopadhyay~\textit{et al.}\cite{Chattopadhyay2006} utilized the same scalar approach as Purcell and estimated the reduced $2 \times 2$ resistance tensor using optical tweezers to trap a sample of swimming \textit{E. coli}.  Recently, Maffeo \textit{et al.}\cite{Maffeo2023} have looked at using rotating nucleic acid double helices as turbines, using electric fields to drive the motion of these molecules. 

The work on helical molecules is at a length scale where the theory of pitch may help guide design parameters for molecular machines. To test these ideas, it is important to determine if the pitch matrix can reproduce previous work on helical systems in general. In this section, we first discuss the pitch matrix properties of a single microhelix using the hydrodynamic model developed in Sec.~\ref{subsec:triangulated_surface_method} to compute the $6 \times 6$ resistance tensor. In the following section, we apply this work to three primary structures of DNA double helices.

\begin{figure}[t]
    \includegraphics[width=0.9\linewidth]{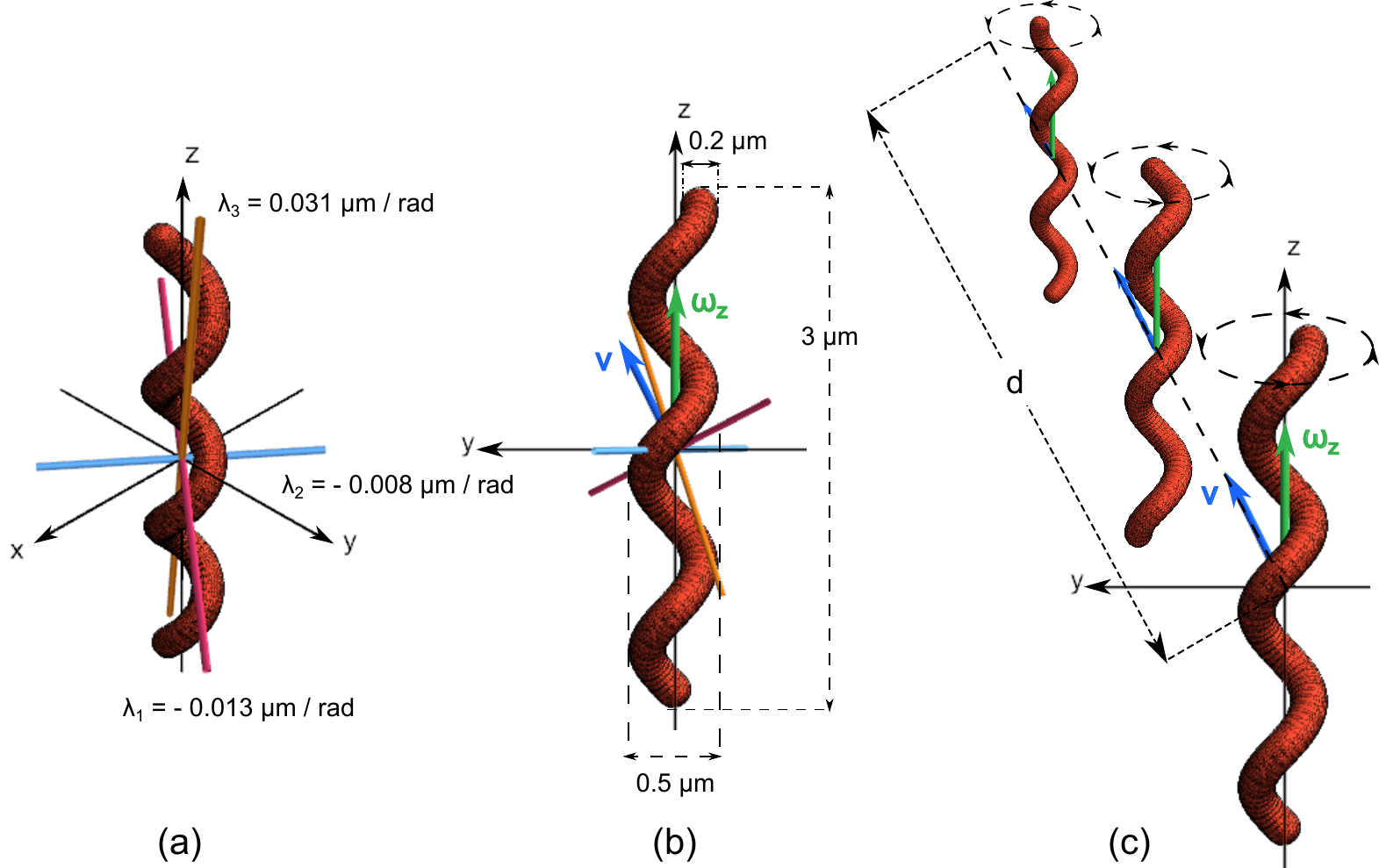}
    \caption{Pitch of a Right-handed Helix: (a) The principal axes of pitch (pink, blue and orange lines) and the three associated moments of pitch for motion around those axes. (b) Design parameters and the velocity vector $\mathbf{v}$ due to rotation around the $z$-axis with an angular velocity $\omega_z$. (c) The distance $\mathrm{d}$ traveled by this helix after rotating around the $z$-axis. This distance can be computed by Eq.~\eqref{eq:projected_distance_of_travel}. \label{fig:helix_triangles}} \hypertarget{fig:helix_triangles}{}
\end{figure}

To test the pitch matrix properties for a simple helix, we constructed a 3~$\mathrm{\mu m}$ right-handed helix with an outside diameter of 0.5 $\mathrm{\mu m}$ and a thickness of 0.2 $\mathrm{\mu m}$ using spherical beads. The helix was constructed through a procedure outlined in the Supporting Information of Ref.~\onlinecite{Duraes2021}, where the centers of consecutive beads are 0.049 $\mathrm{\mu m}$ apart. To triangulate the surface mesh (see Fig. \ref{fig:helix_triangles}) and compute the resistance tensor and pitch matrix for the helix, we used the MSMS algorithm\cite{Sanner96} with a probe radius large enough (0.2 $\mathrm{\mu m}$) to smooth the helical surface. Figure~\ref{fig:helix_triangles} presents the principal axes of pitch and the associated moments of pitch for motion around these axes. None of the principal axes of pitch lie along the long axis ($z$-axis) of the helix. Therefore, a rotation around the helical $z$-axis will result in translational drift, whose direction is indicated by the vector $\mathbf{v}$ in Fig.~\ref{fig:helix_triangles}\protect\hyperlink{fig:helix_triangles}{(b)}. 

Using the pitch matrix computed at the center of pitch, we can predict the resulting motion for this helix as it rotates around the $z$-axis. With an angular velocity $\bm{\omega} = (0,0,1) \mathrm{~rad~s}^{-1}$  and using Eq.~\eqref{eq:pitch_matrix_generalization}, the resulting translational velocity for this helix is $\mathbf{v} = (0.00970, 0.0120, 0.0242) \mathrm{~\mu m~s}^{-1}$. The translational motion of the helix will be along the vector $\mathbf{v}$, and the projected distance of travel, 
\begin{equation}
\label{eq:projected_distance_of_travel}
 \mathrm{d} = |\mathbf{v}| \times t = \frac{|\mathbf{v}|}{|\bm{\omega}|} \times |\bm{\omega}|\,t 
\end{equation}
where $t$ is the total time.

From Chasles' theorem,\cite{Bottema1990} which states that rigid body motion can be decomposed into rotation along an axis and translation parallel to that axis (a screw displacement), the term $\displaystyle{|\mathbf{v}| / |\bm{\omega}|}$ in Eq.~\eqref{eq:projected_distance_of_travel} may be interpreted as the pitch projected along the vector $\mathbf{v}$. For the helix in Fig.~\ref{fig:helix_triangles}, $\displaystyle{|\mathbf{v}| / |\bm{\omega}|} = 0.0287 \mathrm{~\mu m~rad}^{-1}$, and thus it will travel a distance $\mathrm{d} = 180$ nm after one complete revolution around the $z$-axis. Because the helix is moving through a fluid, rather than a solid, the distance $\mathrm{d}$ is smaller than the designed pitch of the helix, which is $1~\mathrm{\mu m}$ per turn.  (As in physical screws, the designed pitch of a helix will only be equivalent to the travel from one rotation when the helix is advancing through a solid substrate.) 

The constructed helix in Fig.~\ref{fig:helix_triangles} is similar to the microhelices propelled with a magnetic field by Patil \textit{et al}.\cite{Patil2021} We note that this group observed the microhelices drifting and estimated an experimental projected  pitch of 250 nm.   In Sec.~\ref{B-sec:helices} of the supplementary material,  we also provide data on three helices which approximate those in the Patil \textit{et al.} experiments\cite{Patil2021} and find projected pitch values from 138--280~nm. 

To make a direct comparison to experiments, we can use the scalar pitch coefficient, a rotational invariant defined in Eq.~\eqref{eq:pitch_coefficient}, which includes contributions from all three pitch axes. In the helix in Fig. \ref{fig:helix_triangles}, the scalar pitch coefficient is calculated to be 125~nm. For helices with flagellar widths ranging from 0.1--0.25~$\mu$m, we find scalar pitch coefficients from 95.5--194~$\mu$m. We also note that the scalar pitch coefficients are all $\sim$70\% of the largest of the three moments of pitch, so we can infer that the helix tends to align to the axis of pitch associated with the largest of the three moments.  Drifting was also observed by Ceylan \textit{et al.}\cite{Ceylan2019} in their experiments with helical microswimmers.

\subsubsection{Double Helices: A-, B-, and Z-DNA\label{subsec:double_helices_DNA}}
To study a biologically relevant set of helices, we analyzed molecular structures representing the A, B, and Z forms of DNA. 
The A-DNA sample is a dodecamer with 3 consecutive CpG steps (PDB code 5MVK),\cite{Hardwick2017} the B-DNA sample is a Dickerson-Drew dodecamer (PDB code 4C64),\cite{Lercher2014} and the Z-DNA sample is also a dodecamer (PDB code 4OCB).\cite{Luo2014} These three DNA structures are all derived from experimental crystal structures.

To triangulate the surface of the DNA samples, we represented the atoms as spheres with appropriate van der Waals radii and used the MSMS algorithm\cite{Sanner96} with a probe radius of 1.41~\AA{} to mimic the surrounding water molecules.\cite{Ben-Naim1972} We computed the resistance tensor employing the triangulated surface method described in Sec.~\ref{subsec:triangulated_surface_method}.

\begin{figure}[t]
    \centering
    \includegraphics[width=\linewidth]{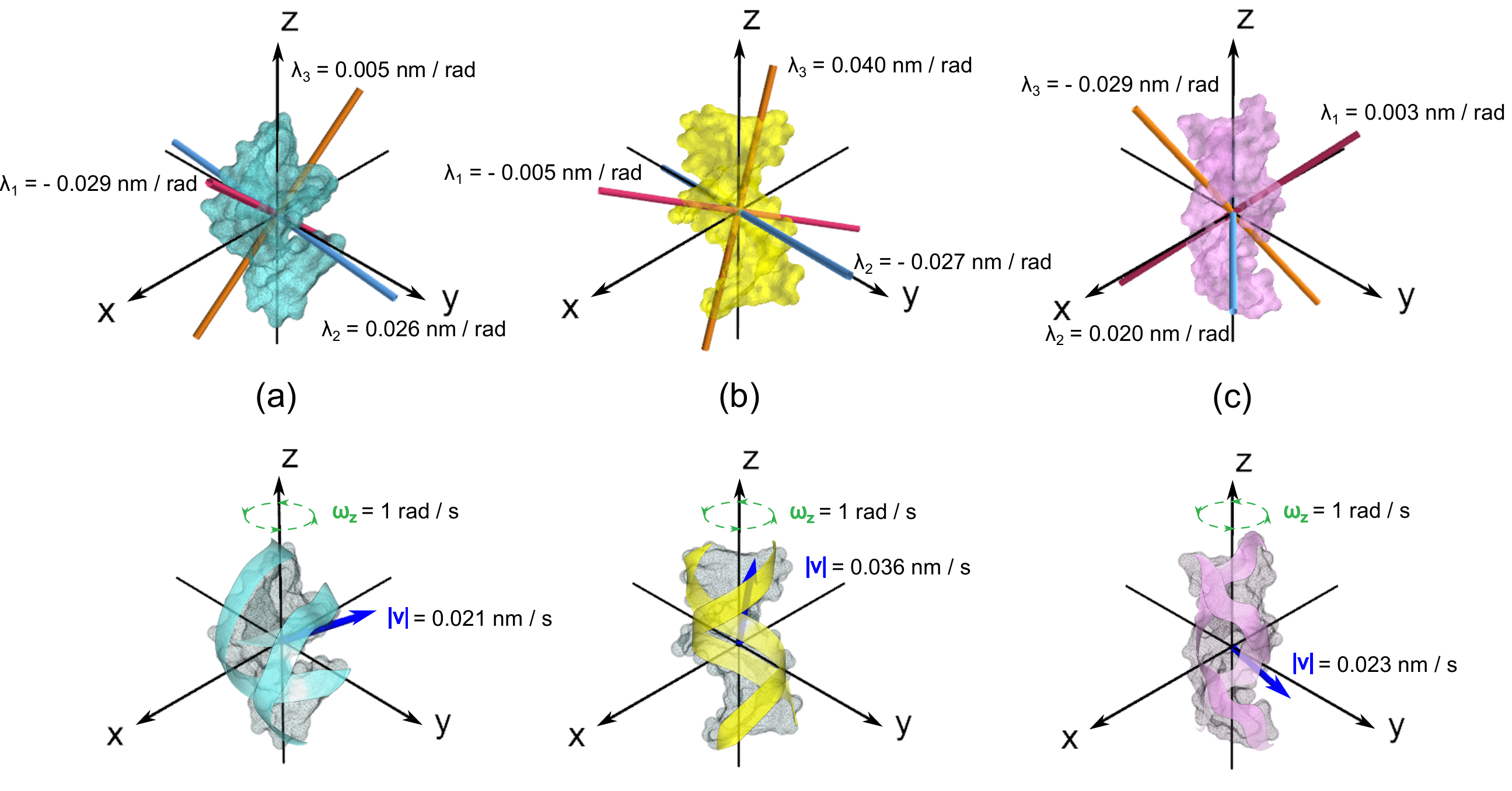}
    \caption{The three DNA samples: (panel~a) A-DNA, (panel~b) B-DNA and (panel~c)~Z-DNA. The principal axes of pitch and their moments of pitch are in the upper panels. The ribbon diagram and the resulting velocity vector caused by a rotation around the $z$-axis are in the lower panels. The ribbon diagram denotes the sugar-phosphate backbones (helices) of the DNA and were generated using the CCP4mg.\cite{McNicholas2011}}
    \label{fig:DNA_triangles}
\end{figure}

In Fig.~\ref{fig:DNA_triangles}, the upper panels display the pitch axes along with the associated moments of pitch for the three DNA samples and the lower panels, the resulting translational velocity due to a rotation around the $z$-axis ($\omega_z = 1 \mathrm{~rad~s}^{-1}$). The three DNA samples manifest translational drift coupled to the rotation around the $z$-axis. From Eq.~\eqref{eq:projected_distance_of_travel}, we can compute the pitches projected along the vector $\mathbf{v}$ displayed in the lower panels, $\displaystyle{\frac{|\mathbf{v}|}{|\bm{\omega}|}\times2\pi=}$ 0.13~nm~(A-DNA), 0.23~nm~(B-DNA) and 0.14~nm~(Z-DNA). These values are smaller than the average structural pitch (per turn)  associated with DNA which are: 2.82~nm (A-DNA), 3.38~nm (B-DNA), and 4.50~nm (Z-DNA).\cite{Saenger1984}  Because the DNA samples are moving in a fluid (and not through a solid substrate), we expect the moments of pitch to be significantly smaller than the structural pitch, since the fluid slips along the surface of the DNA molecules.        

In a stationary liquid, using the Einstein relation between mean square displacement and the translational diffusion coefficient $D_\mathrm{tt}$,\cite{Einstein1956}
\begin{equation} \label{eq:mean_squared_displacement_liquid}
     \langle\, \left(\delta r(t)\right)^2 \,\rangle = 6\, D_\mathrm{tt} \, t\;,
\end{equation}
we can estimate when the translation-rotation coupling will overcome diffusion. This will happen when the ratio of the distance squared in Eq.~\eqref{eq:projected_distance_of_travel} to the mean square displacement in Eq.~\eqref{eq:mean_squared_displacement_liquid} is greater than~1,
\begin{equation} \label{eq:condition_ratio_coupling_diffusion}
     |\bm{\omega}|^2\,t > 6\, D_\mathrm{tt} \,\left(\frac{|P|}{2\,\pi}\right)^{-2}
\end{equation}
where we substituted the term $\displaystyle{|\mathbf{v}| / |\bm{\omega}|}$ in Eq.~\eqref{eq:projected_distance_of_travel} for the scalar pitch coefficient $\displaystyle{|P| / 2\,\pi}$ in Eq.~\eqref{eq:pitch_coefficient}. The diffusion coefficient is calculated as $\displaystyle{D_\mathrm{tt}\!=\!\frac{1}{3} \mathrm{Tr}~\mathsf{D}_\mathrm{CD}^\mathrm{tt}}\,$,\cite{Garcia1980,Garcia1999} where the matrix $\displaystyle{\mathsf{D}_\mathrm{CD}^\mathrm{tt}\!=\left(k_B\,T\right)\,{\mu}_{\mathrm{CD}}^\mathrm{tt}}$ is defined in Eq.~\eqref{eq:mobility_tensor} and CD stands for the center of diffusion (see~Sec.~\ref{subsec:center_of_pitch}).  The term $|\bm{\omega}|^2 \,t$ is a threshold value which can aid in the design of propulsion experiments.    

Table~\ref{tab:DNA_ratio_condition} shows the diffusion coefficients for the DNA samples and the angular velocity conditions for when the translation-rotation coupling overcomes translational diffusion. The translational diffusion coefficients are computed in dilute water solutions at 298.15~K and $\eta~\!=~\!0.89$~mPa$\cdot$s.\cite{CRC_Handbook2022}
From the angular velocity conditions in Table~\ref{tab:DNA_ratio_condition}, a 11.6-day experiment requires angular velocities, $|\bm{\omega}|$, that exceed $1.4 \times 10^3~\mathrm{rad~\mathrm{s}^{-1}}$ (A-DNA), $1.1 \times 10^3~\mathrm{rad~\mathrm{s}^{-1}}$ (B-DNA), and $1.6 \times10^3~\mathrm{rad~\mathrm{s}^{-1}}$ (Z-DNA) to overcome translational diffusion. In longer experiments, as long as rotations are continuous, translation-rotation coupling can overcome diffusion with smaller angular velocities.

\begin{table*}
\caption{Translational diffusion coefficients $D_\mathrm{tt}$ and scalar pitch coefficients $|P|/{2\,\pi}$ (Eq.~\eqref{eq:pitch_coefficient}) for dodecamers of the three helical forms of DNA. The translational diffusion coefficients are based on dilute water solutions at 298.15 K with a viscosity of 0.89~mPa$\cdot$s.\cite{CRC_Handbook2022}
Conditions when translation-rotation coupling overtakes translational diffusion are also provided (Eq.~\eqref{eq:condition_ratio_coupling_diffusion}). \label{tab:DNA_ratio_condition}}
\begin{tabular*}{\linewidth}{@{\extracolsep{\fill}}cccc}
\hline 
Structure (PDB code) & $D_\mathrm{tt}$ $\left(10^{8}~\mathrm{nm}^{2}~\mathrm{s}^{-1}\right)$ & $|P|/{2\,\pi}$ $\left(\mathrm{nm}~\mathrm{rad}^{-1}\right)$ & $|\bm{\omega}|^{2}\times t$ $\left(10^{12}~\mathrm{rad}^{2}~\mathrm{s}^{-1}\right)$  \\ \hline
A-DNA (5MVK) & 1.70 & 0.023 & $> 1.9$ \\
B-DNA (4C64) & 1.65 & 0.028 & $> 1.3$ \\
Z-DNA (4OCB) & 1.66 & 0.020 & $> 2.5$ \\
\hline \hline
\end{tabular*}
\end{table*}

In comparison with the DNA samples in Table~\ref{tab:DNA_ratio_condition}, the $3~\mu \mathrm{m}$ single helix in Sec.~\ref{subsec:helices} has $D_\mathrm{tt}=4.39 \times 10^{5}~\mathrm{nm}^{2}~\mathrm{s}^{-1}$ when suspended in the same dilute water conditions, with a pitch coefficient of $|P|/{2\,\pi}=19.9~\mathrm{nm}~\mathrm{rad}^{-1}$. The $|\bm{\omega}|^2 \,t$ threshold value points to the minimum frequency of rotation $|\omega|$ that the helix must have to overcome translational diffusion. This parameter, $|\bm{\omega}|^{2} t > 6.65\times10^{3}~\mathrm{rad}^{2}~\mathrm{s}^{-1}$, is also dependent on the time scale $(t)$ for the experiment. In a 100-second experiment, the helix will overcome diffusion when its frequency is held constant at a minimum of 1.30~Hz. For the same helix, in a 10-second experiment, the helix will overcome diffusion when its frequency is held constant at 4.10~Hz.      
In an experiment with similar helices and solvent conditions, Patil \textit{et al}.\cite{Patil2021} applied a rotating magnetic field with frequencies in the range 5--15~Hz to propel their microscopic helices, which are well above the predicted minimum threshold frequencies to observe propulsion. Patil \textit{et al}.\cite{Patil2021} also reported that their helices could overcome diffusion when the rotation frequency was 2~Hz.

\subsection{Achiral Swimmers\label{subsec:achiral_swimmers}}

In Sec.~\ref{subsec:pitch_properties}, we showed that achiral objects can be divided into two groups by their moments of pitch. The first group consists of achiral objects for which \textit{all} moments of pitch are zero, i.e., the pitch matrix is a null matrix. As a result, objects in this group exhibit no translation-rotation coupling and rotation will not produce displacement. Examples of these achiral non-swimmers are well-known; \textit{e.g.}, spheres, ellipsoids, tetrahedra, and cubes. Interested readers are encouraged to consult  Sec.~\ref{B-sec:non_swimmers_achiral_swimmers} of the supplementary material  for more details.

The second group consists of achiral objects with a special symmetry, where one moment of pitch is zero, while the other two have the same magnitude, but opposite signs. For these objects, the pitch matrix is non-zero and translation-rotation coupling persists. These objects have previously been called \textit{achiral swimmers} because they produce displacements due to rotation. Figure~\ref{fig:achiral} displays an achiral microswimmer: an arrangement of three beads that was experimentally tested by Cheang \textit{et al.}\cite{Fu2014} In their work, Cheang~\textit{et al.} reported that a non-vanishing (rt) block of the mobility tensor is required for swimming and, from the symmetry investigations conducted in Ref.~\onlinecite{Brenner1964}, concluded that \textit{achiral swimmers} are real. 




To triangulate the surface of the three-beads arrangement in Fig.~\ref{fig:achiral}, we used the MSMS algorithm\cite{Sanner96} with a probe radius of $10^{-6}~\mu \mathrm{m}$ and a triangulation density of 20~$\mathrm{vertices/}\mu\mathrm{m}^2$. In Fig.~\ref{fig:achiral}, we also found the principal axes of pitch and their associated moments employing the Bead Model developed in Ref.~\onlinecite{Duraes2021} in addition to the boundary element approach developed in this work (Sec.~\ref{subsec:triangulated_surface_method}). As expected, with both methods of finding the resistance tensor, one moment of pitch is zero and the other two have the same magnitude with a flipped sign.

Utilizing the bead and the boundary element models, Table~\ref{tab:three_beads_ratio_condition} presents the translational diffusion and the scalar pitch coefficient for the three-bead achiral swimmer along with the angular velocity condition for when the translation-rotation coupling surpasses diffusion. Translational diffusion coefficients were computed at 298.15 K and $\eta=1.0$~mPa$\cdot$s, reproducing the experimental conditions of Cheang \textit{et al.}'s work\cite{Fu2014} for a \ce{NaCl} solution. In the first second of an experiment, the translation-rotation coupling of the immersed three-bead swimmer will overtake diffusion when the angular velocity, $|\bm{\omega}|$, exceeds 7.6~$\mathrm{rad~\mathrm{s}^{-1}}$ (Bead Model) and 8.0~$\mathrm{rad~\mathrm{s}^{-1}}$ (Boundary Element Model). These are equivalent to rotational frequencies that exceed 1.3 Hz, and are comparable to the rotating magnetic field frequencies of 1--8 Hz applied by Cheang \textit{et al.}\cite{Fu2014} to propel their swimmers. The scalar pitch values in Table~\ref{tab:three_beads_ratio_condition} and the moments of pitch in Fig.~\ref{fig:achiral} can be used to generate similar swimming speeds reported in Fig. 3(b) of Ref.~\onlinecite{Fu2014}.

Translational drift is expected when the axes of rotation are not the principal axes of pitch. This translational drift can be seen clearly in the supporting videos\cite{Fu2014} displaying the motion of the three-beads swimmers. Hermans \textit{et al.}\cite{Hermans2015} also reported translational drift in an experiment with a rotating achiral swimmer. In a Taylor--Couette device, Hermans \textit{et al.}'s achiral swimmer had one orbital radius when rotating clockwise and another when rotating counterclockwise.

\begin{table*}
\caption{Translational diffusion coefficients and scalar pitch coefficients for the achiral three-beads swimmer in Fig.~\ref{fig:achiral}. The translational diffusion coefficients are calculated at 298.15 K with a viscosity of 1.0~mPa$\cdot$s to mimic the conditions in Ref. \onlinecite{Fu2014}. Conditions when translation-rotation coupling overtakes translational diffusion are also provided (Eq.~\eqref{eq:condition_ratio_coupling_diffusion}). \label{tab:three_beads_ratio_condition}}

\begin{tabular*}{\linewidth}{@{\extracolsep{\fill}}cccc}
\hline 
Hydrodynamic Model & $D_\mathrm{tt}$ $\left(10^{4}~\mathrm{nm}^{2}~\mathrm{s}^{-1}\right)$ & $|P|/{2\,\pi}$ $\left(\mathrm{nm}~\mathrm{rad}^{-1}\right)$ & $|\bm{\omega}|^{2}\times t$ $\left(\mathrm{rad}^{2}~\mathrm{s}^{-1}\right)$  \\ \hline
Bead (Ref.~\onlinecite{Duraes2021}) & 6.07 & 79 & $> 58$ \\
Boundary Element (Sec.~\ref{subsec:triangulated_surface_method}) & 5.80 & 74 & $> 64$ \\
\hline \hline
\end{tabular*}
\end{table*}

\subsection{Lord Kelvin's Isotropic Helicoid \label{subsec:isotropic_helicoid}}
An \textit{isotropic helicoid} is an object for which the blocks of the resistance tensor are isotropic at the center of resistance (CR). That is, the four blocks may be written:\cite{Brenner1964} 
\begin{align} \label{eq:scalar_helicoid}
\Xi^{\mathrm{tt}} & =\xi^{\mathrm{tt}}\,\mathbf{I} \nonumber \\  
\Xi^{\mathrm{rt}}_{\mathrm{CR}} &=\Xi^{\mathrm{tr}}_{\mathrm{CR}}=\xi^{\mathrm{tr}}\,\mathbf{I} \\
\Xi^{\mathrm{rr}}_\mathrm{CR} & =\xi^{\mathrm{rr}}\,\mathbf{I} \nonumber
\end{align}
where $\xi^{\mathrm{tt}}$, $\xi^{\mathrm{tr}}$ and $\xi^{\mathrm{rr}}$ are scalars and $\mathbf{I}$ is the $3 \times 3$ identity matrix. The only difference between these objects and spherically isotropic bodies (\textit{i.e.}, spheres, cubes and tetrahedra) is that helicoids have non-zero rotation-translation coupling
$\displaystyle{(\xi^{\mathrm{tr}} \neq 0)}$.\cite{Brenner1964}

\begin{figure}
    \includegraphics[width=\linewidth]{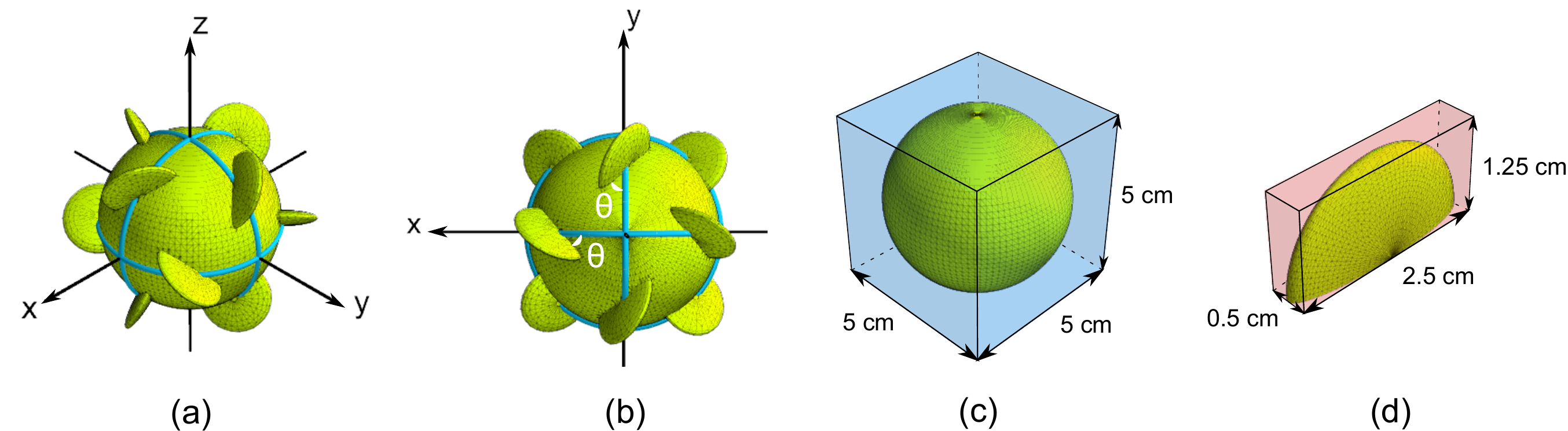}
    \caption{Generalization of Lord Kelvin's Isotropic Helicoid. (a) Vanes are placed midway between the intersection points of three circles around a sphere. (b) The vanes are tilted at an angle $\theta$ relative to the circles, and this angle defines the handedness of the isotropic helicoid. The dimensions of (c) the sphere and (d) the semi-oblate vanes used in our examples. \label{fig:isotropic_helicoids_construction}}
    \hypertarget{fig:isotropic_helicoids_construction}{}
\end{figure}

In 1871, Sir William Thomson (widely known as Lord Kelvin) proposed one design for an isotropic helicoid using a sphere with 12 projecting vanes arranged in a systematic way.\cite{Thomson1871} A generalization of his approach is shown in Fig.~\ref{fig:isotropic_helicoids_construction} and is described below:\cite{Frost2021,Gustavsson2016}

\begin{enumerate}
    \item Center a sphere at the origin, and locate three circles at the intersections of the sphere with the $xy$-, $yz$- and $xz$-planes (Fig.~\ref{fig:isotropic_helicoids_construction}\protect\hyperlink{fig:isotropic_helicoids_construction}{(a)}).
    \item Using the six intersection points of the three circles, place the centers of semi-oblate vanes midway between these intersection points. In the end, there will be 4 semi-oblate vanes per circle (Fig.~\ref{fig:isotropic_helicoids_construction}\protect\hyperlink{fig:isotropic_helicoids_construction}{(b)}).
    \item The orientation angles $\theta$ of the vanes are related to their positions in the circles and defines the handedness of the isotropic helicoid. The isotropic helicoid is right-handed for $0^\circ<\theta<90^\circ$ and left-handed for $-90^\circ<\theta<0^\circ$. This definition comes from the sign of the angular velocity $\omega$ in Eq.~\eqref{eq:terminal_velocity_angular_velocity} and the direction of rotation of a right-handed screw.   
\end{enumerate}

\begin{figure}
    \includegraphics[width=\linewidth]{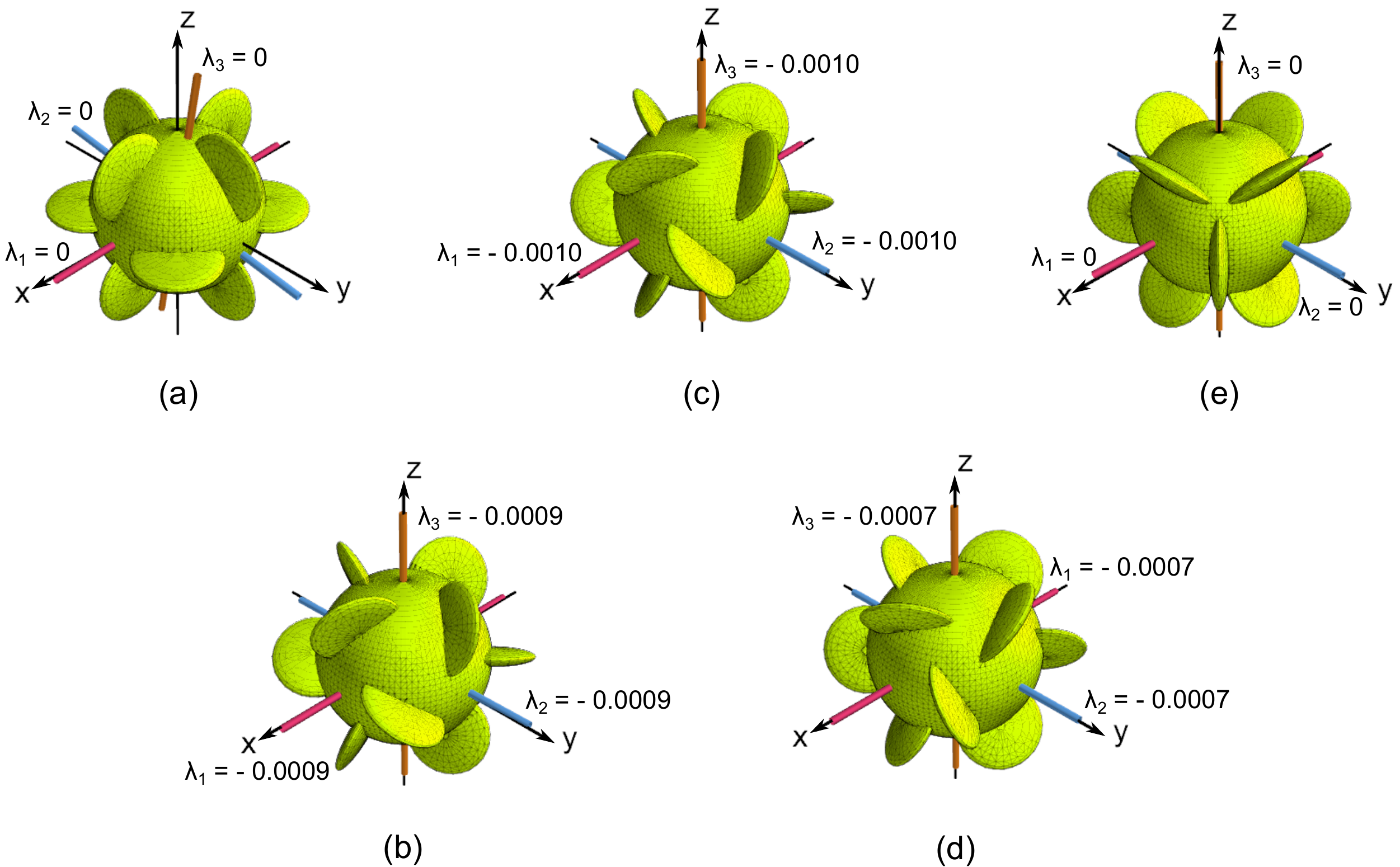}
    \caption{ Moments of pitch (in $\mathrm{cm/rad}$) for isotropic helicoids with different vane angles as described in Fig.~\ref{fig:isotropic_helicoids_construction}. When the vanes are at $0^\circ \mathrm{~or~} -90^\circ$ (panels (a) and (e)), the objects are non-swimmers and do not couple rotation to translation. Pitch axes and moments of pitch indicate that all other vane angles generate non-zero rotation-translation coupling. Here, we show helicoid moments of pitch when the vanes are at $-30^\circ$, $-45^\circ$, or $-60^\circ$ (panels (b), (c) and (d)). }
    \label{fig:kelvin}
    \hypertarget{fig:kelvin}{}
\end{figure}

We constructed three left-handed isotropic helicoids $(\theta = -30^\circ, -45^\circ, -60^\circ)$ and two spherically isotropic bodies $(\theta = 0^\circ, -90^\circ)$ utilizing the procedure in Ref.~\onlinecite{Frost2021}. Each of these isotropic objects are composed of a sphere with 12 semi-oblate vanes whose dimensions are displayed in Fig.~\ref{fig:isotropic_helicoids_construction}\protect\hyperlink{fig:isotropic_helicoids_construction}{(c)} and \protect\hyperlink{fig:isotropic_helicoids_construction}{(d)}. Triangulation of these objects was performed in OpenSCAD,\cite{OpenSCAD} and we analyzed them employing the boundary element method described in Sec.~\ref{subsec:triangulated_surface_method} with a 120-point quadrature developed by Xiao and Gimbutas.\cite{Xiao2010} This quadrature exactly integrates polynomials of degree 25 and its points and weights are available in Quadpy.\cite{Quadpy_github}  Figure~\ref{fig:kelvin} shows the pitch axes and the moments of pitch for these isotropic objects. 

For all five objects in Fig.~\ref{fig:kelvin}, both the center of resistance and the center of pitch are located at the origin. For the two spherically isotropic bodies, $\displaystyle{\Xi^{\mathrm{rt}}_{\mathrm{CR}}=\Xi^{\mathrm{tr}}_{\mathrm{CR}}=0}$. From Eq.~\eqref{eq:pitch_matrix_resistance_tensor}, this implies that the pitch matrix itself is zero and no translation-rotation coupling is possible. The isotropic helicoids $(\theta = -30^\circ, -45^\circ, -60^\circ)$ have non-zero moments of pitch, but our calculations indicate that these are quite small.  From the definition of the pitch matrix, these small moments of pitch are related to small rotation-translation coupling values ($\xi^{\mathrm{tr}}$), and we can use these moments of pitch to demonstrate rotation-translation coupling of isotropic helicoids. For example, in an experiment where an isotropic helicoid from Fig. \ref{fig:kelvin} with $\theta = -45^\circ$ is suspended in a viscous fluid and a rotation frequency of 10~Hz is imposed along one of the pitch axes, the helicoid will move 6.3 cm in a 100~s observation time. On the other hand, the spherically isotropic bodies $(\theta = 0^\circ, -90^\circ)$ will not move at all, since there is no rotation-translation coupling.  In the supplementary material (Sec.~\ref{B-sec:isotropic_helicoids_sm}),  we compute pitch axes and moments of pitch for a related set of left-handed isotropic helicoids, and we show the pitch coefficients as a function of the vane angle $(\theta)$.

Table~\ref{tab:isotropic_helicoids_blocks_resistance_tensors} presents the scalar values associated with the blocks of the resistance tensor (Eq.~\eqref{eq:scalar_helicoid}) for the isotropic objects in Fig.~\ref{fig:kelvin}. These scalar values were computed in silicon oil with $\eta=490$~mPa$\cdot$s~\cite{Collins2021} and employing the triangulation and boundary element method described above. For $\theta=-30^\circ$, $-45^\circ$ and $-60^\circ$, the scalar values are related to the isotropic helicoids where $\xi^{\mathrm{tr}}\neq0$. For $\theta=0^\circ$ and $-90^\circ$, the scalar values are related to the spherically isotropic bodies where $\xi^{\mathrm{tr}}=0$.

\begin{table}[h]
\caption{Scalar parameters for the translational (tt), translation-rotation (tr), and rotational (rr) blocks of the resistance tensor for isotropic helicoids with different vane angles $(\theta)$. The moments of pitch for these objects can be obtained from $-\,\xi^\mathrm{tr} / \xi^\mathrm{tt}\,$. Also shown are the terminal velocities, $\mathrm{v}$, and terminal angular velocities, $\omega$, for these bodies as they fall through a viscous fluid. Using the experimental conditions in Ref.~\onlinecite{Collins2021}, the terminal angular velocities are small, corroborating experimental observations.
\label{tab:isotropic_helicoids_blocks_resistance_tensors}}

\begin{tabular*}{\linewidth}{@{\extracolsep{\fill}}cccccc}
\hline 
\multirow{2}{*}{Angle}  & $\xi^{\mathrm{tt}}$  & $\xi^{\mathrm{tr}}$  & $\xi^{\mathrm{rr}}$ & $\mathrm{v}$ & $\omega$  \\ 
 & $\left(\mathrm{kg/s}\right)$ &  $\left(\mathrm{kg~m/(s~rad)} \right)$ & $\left(\mathrm{kg~m^2/(s~rad^2)}\right)$  & ($\mathrm{cm/s}$) & ($\mathrm{rad/s}$)  \\  \hline 
$\phantom{-}0^\circ$ & 0.284 & 0 & $3.73\times 10^{-4}$ & 46.8 & 0 \\
$-30^\circ$ & 0.283  & $2.65 \times 10^{-6}$& $3.73\times 10^{-4}$  & 47.0 & $-3.34\times 10^{-3}$ \\
$-45^\circ$ & 0.283 & $2.72 \times 10^{-6}$& $3.73\times 10^{-4}$  & 47.0 & $-3.42\times 10^{-3}$ \\
$-60^\circ$ & 0.283  & $1.89 \times 10^{-6}$& $3.73\times 10^{-4}$  & 47.0 & $-2.38\times 10^{-3}$ \\
$-90^\circ$ & 0.283  & 0 & $3.73\times 10^{-4}$ & 47.0 &  0 \\
\hline \hline
\end{tabular*}

\end{table}

 Rotation-translation coupling can also be demonstrated if we allow an isotropic helicoid or spheroid to fall through a quiescent fluid. The force and torque on the body are:\cite{Brenner1964,Collins2021} 
\begin{equation} \label{eq:motion_helicoids_gravitational_field}
\left[\begin{array}{c}
\mathbf{f}\\
\bm{\tau}
\end{array}\right]= - \, \left[\begin{array}{cc}
\Xi^{\mathrm{tt}} & \Xi^{\mathrm{rt}}_{\mathrm{CR}}\\
\Xi^{\mathrm{tr}}_{\mathrm{CR}} & \Xi^{\mathrm{rr}}_{\mathrm{CR}}
\end{array}\right] \, \left[\begin{array}{c}
\mathbf{v}\\
\bm{\omega}
\end{array}\right]\,+\,\left(m_b \,- \,m_{\!f}\right)\, \left[\begin{array}{c}
\mathbf{g}\\
\mathbf{0}
\end{array}\right] \quad,
\end{equation}
where $m_b$ and $m_{\!f}$ are the masses of the body and the displaced fluid, respectively, $\mathbf{g}$ is the gravitational vector field and $\mathbf{0}$ is the null vector. The term $m_b\,\mathbf{g}$ represents the gravitational force and the term $-m_{\!f}\,\mathbf{g}\,$, the force due to buoyancy.

Considering that the body is falling along a single axis, and using the isotropy definitions in Eq.~\eqref{eq:scalar_helicoid}, we can compute the terminal velocity and angular velocity,\cite{Brenner1964}
\begin{equation} \label{eq:terminal_velocity_angular_velocity}
    \mathrm{v} = \frac{\xi^{\mathrm{rr}} \left(m_b \, - \, m_{\!f}\right)g}{\xi^\mathrm{tt}\,\xi^\mathrm{rr}\,-\,\left(\xi^\mathrm{tr}\right)^2} \quad \mathrm{and} \quad \omega = \frac{-\,\xi^{\mathrm{tr}}\left(m_b\,-\,m_{\!f}\right)g}{\xi^\mathrm{tt}\,\xi^\mathrm{rr}\,-\,\left(\xi^\mathrm{tr}\right)^2}\quad.
\end{equation}

For chiral objects, we know from our previous work in Ref.~\onlinecite{Duraes2021} that only $\xi^\mathrm{tr}$ will reverse sign. This implies that $\mathrm{v}$ in Eq.~\eqref{eq:terminal_velocity_angular_velocity} is the same for both a helicoid and its mirror image, but the angular velocity $\omega$ will flip sign for the mirror image (enantiomeric) version of the object. 

For the angular velocity in Eq.~\eqref{eq:terminal_velocity_angular_velocity}, Brenner\cite{Brenner1964} employed the standard definition of $\omega > 0$ for counterclockwise rotations, following the right-handed rule.\cite{Bhandari2010} This implies $\xi^\mathrm{tr} < 0$ for right-handed isotropic helicoids, since the scalars $\xi^{\mathrm{tt}}\,$, $\xi^{\mathrm{rr}}$ and $\left[\xi^\mathrm{tt}\,\xi^\mathrm{rr}\,-\,\left(\xi^\mathrm{tr}\right)^2\right]$ are always positive.\cite{Brenner1964} We note that this definition is the opposite of the one employed by Gustavsson and Biferale\cite{Gustavsson2016} and Collins \textit{et al.}\cite{Collins2021}  

Table~\ref{tab:isotropic_helicoids_blocks_resistance_tensors} also provides the terminal velocities and terminal angular velocities for the bodies in Fig.~\ref{fig:kelvin}. To compute these values, we used $g = 9.81~\mathrm{m/s}^{2}$ and the experimental conditions in Ref.~\onlinecite{Collins2021}, \textit{i.e.}, body density $\rho_b=1.16~\mathrm{g/cm}^3$ and fluid density $\rho_{\!f}=0.98~\mathrm{g/cm}^3$ (silicon oil). The mass contributions to the gravitational and buoyant forces can be calculated from the volume of the rigid body,  $V_b = 75.267~\mathrm{cm}^3$,  which is the combined volume of the sphere and the 12 semi-oblate vanes in Fig.~\ref{fig:isotropic_helicoids_construction}. The spherically isotropic bodies will not manifest angular velocities because their scalar $\xi^{\mathrm{tr}}=0$. We predict that the left-handed isotropic helicoids will manifest an angular velocity on the order of $10^{-3}~\mathrm{rad/s}$ in a clockwise direction.

The scalar friction values, $\xi^{\mathrm{tt}}$, $\xi^{\mathrm{tr}}$ and $\xi^{\mathrm{rr}}$ scale as $\eta\,L$, $\eta\,L^2$ and $\eta\,L^3$, respectively, where $L$ is the length of the rigid body. Since the volume $V_b$ scales with $L^3$, we conclude that the terminal velocity $\mathrm{v}$ and angular velocity $\omega$ in Eq.~\eqref{eq:terminal_velocity_angular_velocity} will scale as $L^2 / \eta$ and $L / \eta$, respectively. To compare with Collins \textit{et al.}'s helicoid,\cite{Collins2021} we can scale the size of our isotropic helicoid $(\theta=-45^\circ)$ by 0.348 in the same fluid, and obtain a scaled  $\mathrm{v}^{\prime}=\left(0.348\right)^2 \,\mathrm{v} = 5.69~\mathrm{cm/s}$, and a scaled $\omega^{\prime}=0.348 ~\omega = -1.19\times10^{-3}~\mathrm{rad/s}$. Collins \textit{et al.}\cite{Collins2021} reported $\mathrm{v}^{\prime}=4.74~\mathrm{cm/s}$ and $\omega^{\prime}=-0.003~\mathrm{rad/s}$. Since $\omega^{\prime}$ from Collins \textit{et al.} has 1\% uncertainty, our calculation reveals that more sensitive instruments would be required to measure rotation-translation coupling of Collins \textit{et al.}'s helicoid.      



\section{Conclusion}
We have presented a general theory for the pitch of objects which are interacting with a fluid medium at low Reynolds number. The pitch matrix, defined in Eq. \eqref{eq:pitch_matrix_generalization}, is diagonalized to yield three pitch axes along with their associated ``moments of pitch''. The pitch axes and moments arise out of the \textit{geometry} of the objects' surfaces, and they have a number of important properties. First, the symmetry of the object defines the number of degenerate and non-zero  moments of pitch. Second, chiral objects (molecules, helices) couple rotational and translational motion in the fluid, and will move in the opposite direction from their enantiomers (mirror images) under the same rotation. Third, the pitch matrix also provides an explanation for the rotation-translation coupling that allows \textit{achiral} swimmers to migrate when they rotate in a fluid. This theory also helps us to understand the translational drift of rotating helical objects. There are many potential uses of this theory, but the primary interest to chemists is to develop an efficient method for separating enantiomers without costly synthetic pathways currently in use.

One of our primary observations is that chiral objects with a $C_n$ axis of symmetry have two degenerate moments of pitch when $n \ge 3$, and there is no drift for rotations around that axis of symmetry. This appears to be the case for propeller-shaped molecules, and this observation points to a general and efficient design principle.

There are many ways to approximate the hydrodynamic resistance tensor, and we have developed a boundary element method which obeys the symmetry properties of the blocks of the resistance tensor. This was also true of earlier methods that use small beads or atomic spheres to represent the surface of an object or the surface of a molecule, but the method for triangulated surfaces given here is generally applicable to rigid bodies of arbitrary shapes.

Our theory of pitch has been tested against some experiments on microswimmers, and our predictions agree well with experimental observations of translational-rotational coupling.  We also show results for an object of historical curiosity, Lord Kelvin's isotropic helicoid, which can exhibit a small angular velocity as it falls through a fluid. For the collective behavior of a multi-molecule system, consider Ref.~\onlinecite{Duraes2021}, where a competition model was developed to study the separation of chiral molecules in solution. Note that even in racemic mixtures, separation can be achieved under sufficiently large solution vorticities.

There are many potential uses of this theory. Projection of molecular dipoles onto the pitch axis with the largest moment of pitch can help design polarized microwave methods which separate enantiomers through molecular rotation in the fluid (instead of rotating the fluid around the enantiomers). Additionally, this method also now allows us to identify geometries of achiral swimmers from the eigenvalue structure of the pitch matrix. We also now have a firmer understanding of non-axial drift of chiral objects due to projections of angular velocity onto the axes of pitch.

\section{Supplementary Material}
See the supplementary material for additional properties of the pitch matrix and the pitch coefficient, as well as applications of the theory of pitch in isotropic helicoids, achiral swimmers, non-swimmers, and analytically-solvable objects. The supplementary material also develops a relationship between pitch and the moment of inertia for a sphere, and this is used to analyze translation-rotation coupling in spheres rotating in non-Newtonian fluids. An accompanying set of text files provide molecular geometries for the enantiomers and DNA structures, as well as triangulated surfaces for the helix, achiral swimmers, and isotropic helicoids. Code which computes the blocks of the resistance tensor and pitch matrices for these objects is also included. 

\begin{acknowledgments}
  Support for this project was provided by the National Science
  Foundation under Grant No. CHE-1954648. Computational time was provided by the Center for Research Computing (CRC) at the University of Notre Dame.
\end{acknowledgments}

\section*{Data Availability Statement}
The data that support the findings of this study are available within the article and its supplementary material.

\bibliography{main.bib}
\end{document}


\title{Supplementary Material for ``A Theory of Pitch for the Hydrodynamic Properties of Molecules, Helices, and Achiral Swimmers at Low Reynolds Number"}

\author{Anderson D. S. Duraes}
\affiliation{251 Nieuwland Science Hall, \\
  Department of Chemistry \& Biochemistry, \\
  University of Notre Dame, Notre Dame, Indiana 46556, USA}
\author{J. Daniel Gezelter}
\email{gezelter@nd.edu}
\affiliation{251 Nieuwland Science Hall, \\
  Department of Chemistry \& Biochemistry, \\
  University of Notre Dame, Notre Dame, Indiana 46556, USA}

\date{\today}

\begin{abstract}
\vspace{0.7in}
This document provides additional properties of the pitch matrix and the pitch coefficient. It also contains other applications of the theory of pitch in isotropic helicoids, achiral swimmers and non-swimmers. With the boundary element method developed in the main paper, moments and principal axes of the blocks of the resistance tensor are compared with known analytical expressions. These moments and principal axes show good agreement with the analytical values. A relationship between pitch and moment of inertia for a sphere is derived. The behavior of spheres with different radii in non-Newtonian fluids is also analyzed using the theory of pitch.
\end{abstract}

\maketitle

\clearpage
\scshape
\bfseries
\tableofcontents
\normalfont
\clearpage

\section{Proof of the Uniqueness of the Center of Pitch \label{sec:proof_uniqueness_center_of_pitch}}
The center of pitch $\left(\mathbf{p}\right)$ is unique, and this uniqueness can be proven by contradiction. Suppose there is a point $\mathbf{q} \neq \mathbf{p}$ such that ${\displaystyle P_\mathbf{q} = \left(P_\mathbf{q} \right)^T}$. We begin at the center of pitch:
\begin{align}
\label{eq:center_pitch_p_proof_unique}
\frac{P_\mathbf{p}}{2\pi} & =\left(\frac{P_\mathbf{p}}{2\pi}\right)^{T} \\ \intertext{Using Eq.~\eqref{A-eq:pitch_matrix_transform} from the main paper, we find the pitch at the new point $\mathbf{q}$:}
\frac{P_\mathbf{q}}{2\pi} -\mathsf{U}_\mathbf{p} & = \left(\frac{P_\mathbf{q}}{2\pi} -\mathsf{U}_\mathbf{p}\right)^{T} \\ \intertext{Because $\mathsf{U}$ is a skew-symmetric matrix (Eq.~\eqref{A-eq:skew_symmetric_matrix}), its transpose changes sign:} 
\frac{P_\mathbf{q}}{2\pi}-\mathsf{U}_\mathbf{p} & =\left(\frac{P_\mathbf{q}}{2\pi}\right)^{T}+\mathsf{U}_\mathbf{p} \\
\intertext{Rearranging:}
\begin{split}
\frac{P_\mathbf{q}}{2\pi}-\left(\frac{P_\mathbf{q}}{2\pi}\right)^{T} & =2\,\mathsf{U}_\mathbf{p}\\ \\
\implies \mathsf{U}_\mathbf{p} & = \mathbf{0} \\ 
\implies \mathbf{r}_\mathbf{p} & = \left[\begin{array}{c}
x_{\mathbf{qp}}\\
y_{\mathbf{qp}}\\
z_{\mathbf{qp}}
\end{array}\right] =\left[\begin{array}{c}
0\\
0\\
0
\end{array}\right] \\
\implies \mathbf{q} & = \mathbf{p} 
\end{split} 
\end{align}
which is a contradiction. Therefore, the location of the center of pitch is unique for each body. The terms $x_{\mathbf{qp}}$, $y_{\mathbf{qp}}$ and $z_{\mathbf{qp}}$ are the components of the vector between the point $\mathbf{q}$ and the point $\mathbf{p}$.

\section{Pitch Coefficient \label{sec:pitch_coefficient_generalization}}

The pitch coefficient (Eq.~\eqref{A-eq:pitch_coefficient}) can be derived from the term $\displaystyle{\frac{|\mathbf{v}|}{|\bm{\omega}|}}$ in Eq.~\eqref{A-eq:projected_distance_of_travel}. \vspace{0.05in} For an angular velocity vector with equal components, $\displaystyle{\bm{\omega}=(\omega,\,\omega,\,\omega)}$,
\begin{align}
\label{eq:pitch_coefficient_derivation}
\begin{split}
\\
 \frac{|P|}{2 \pi} = \frac{\left|\mathbf{v}\right|}{|\bm{\omega}|} & =  \frac{\displaystyle{\left|\frac{\mathsf{P}}{2\pi}\,\bm{\omega}\right|}}{|\bm{\omega}|}  \\ \\
 & =  \frac{\displaystyle{\sqrt{\left(\frac{\mathsf{P}}{2\pi}\,\bm{\omega}\right)^{\dagger}\left(\frac{\mathsf{P}}{2\pi}\,\bm{\omega}\right)}}}{\sqrt{\bm{\omega}^{\dagger}\, \bm{\omega}}} \\ \\
 & =  \frac{\displaystyle{\sqrt{\omega^2\sum_i \lambda_i^{*} \, \lambda_i^{\phantom{*}}}}}{\sqrt{3\,\omega^2}} \\ \\
 & =  \displaystyle{\sqrt{\frac{1}{3}\,\sum_i \lambda_i^{*} \, \lambda_i^{\phantom{*}}}} 
\end{split} 
\end{align}
\vspace{0.5cm} 

$\!\!\!\!\!\!\!\!\!\!\!$where we have written the pitch matrix in its diagonal form. \vspace{0.05in}Note that at the center of pitch (Sec.~\ref{A-subsec:center_of_pitch}), where the pitch matrix $\displaystyle{\left(\frac{\mathsf{P}}{2\pi}\right)}$ is symmetric, the diagonalization is guaranteed and \vspace{0.05in}the eigenvalues are real numbers,\cite{Riley_Hobson_Bence2006} \textit{i.e.}, $\lambda_i^{*}=\lambda_i^{\phantom{*}}$, recovering Eq.~\eqref{A-eq:pitch_coefficient}.  

\clearpage
\section{Analytical Expressions for the Blocks of the Resistance Tensor \label{sec:analytical_expressions}}

There are geometrical shapes whose analytical expressions for the translational (tt), rotational (rr), translation-rotation (tr) and rotation-translation (rt) blocks of the resistance tensor in Eq.~\eqref{A-eq:resistance_tensor} are known. At the center of resistance (CR)~(Sec.~\ref{A-subsec:center_of_pitch}), these blocks have the form:\cite{Brenner1964}
\begin{equation} \label{eq:geometries_analytical_form}
        \Xi^{\mathrm{tt}} = \left[\begin{array}{ccc}
    \xi^{\mathrm{tt}}_{1} & 0 & 0\\
    0 & \xi^{\mathrm{tt}}_{2} & 0 \\
    0 & 0 & \xi^{\mathrm{tt}}_{3}
    \end{array}\right],\qquad
    \Xi^{\mathrm{rr}}_{\mathrm{CR}} = \left[\begin{array}{ccc}
    \xi^{\mathrm{rr}}_{1} & 0 & 0\\
    0 & \xi^{\mathrm{rr}}_{2} & 0 \\
    0 & 0 & \xi^{\mathrm{rr}}_{3}
    \end{array}\right], \qquad
    \Xi^{\mathrm{tr}}_{\mathrm{CR}} = \Xi^{\mathrm{rt}}_{\mathrm{CR}} = \mathbf{0}
\end{equation}
where $\mathbf{0}$ is the $3 \times 3$ null matrix, and the eigenvalues $\left(\xi^{\mathrm{tt}}_i\right)$ and $\left(\xi^{\mathrm{rr}}_i\right)$ are, respectively, the moments of the translational (tt) and the moments of the rotational (rr) blocks.

In a fluid with viscosity~$\eta$, we provide examples of objects and their analytical expressions for the blocks of the resistance tensor in the form of Eq.~\eqref{eq:geometries_analytical_form}. For these objects, we use the parameters of an ellipsoid whose semi-axes have lengths $a$, $b$ and $c$, and whose equation in Cartesian coordinates is:\cite{Riley_Hobson_Bence2006} 
\begin{equation} \label{eq:ellipsoid_equation}
    \frac{x^2}{a^2} + \frac{y^2}{b^2} + \frac{z^2}{c^2} = 1  
\end{equation}

\vspace{0.5cm}
\begin{enumerate}
\item \textbf{Sphere} $\left(a=b=c=r\right)$:\cite{Torre:1983lr,Duraes2021,Berg1993,Landau1987} \label{itm:sphere}\\
    \begin{subequations}
    \label{eq:sphere_blocks}
    \begin{align} 
    \displaystyle{\xi^{\mathrm{tt}}_1} & = \xi^{\mathrm{tt}}_2 = \xi^{\mathrm{tt}}_3 = \displaystyle{6\pi\eta\, r} \label{eq:sphere_tt_block} \\  
    \xi^{\mathrm{rr}}_1 & = \xi^{\mathrm{rr}}_2 = \xi^{\mathrm{rr}}_3 = \displaystyle{8\pi\eta\, r^3} \label{eq:sphere_rr_block}
     \end{align}
     \end{subequations}
     
     \vspace{0.5cm}
\item \textbf{Oblate Spheroid} $\left(a < b=c\right)$:\cite{Perrin1934,Koenig1975} \label{itm:oblate_spheroid} \\
    \begin{subequations}
    \label{eq:oblate_spheroid_blocks}
    \begin{align} 
    \xi^{\mathrm{tt}}_1 & = \displaystyle{16\pi\eta\,\frac{a^2 \,-\, b^2}{\left(2a^2-b^2\right)S - 2a}} & \xi^{\mathrm{tt}}_2 & = \xi^{\mathrm{tt}}_3 = \displaystyle{32\pi\eta\,\frac{a^2 \,-\, b^2}{\left(2a^2-3b^2\right)S + 2a}} \label{eq:oblate_spheroid_tt_block} \\
    \xi^{\mathrm{rr}}_1 & = \displaystyle{\frac{32\pi\eta}{3}\, \frac{\left(a^2-b^2\right)b^2}{2a\,-\,Sb^2}} & \xi^{\mathrm{rr}}_2 & = \xi^{\mathrm{rr}}_3 = \displaystyle{\frac{32\pi\eta}{3}\, \frac{a^4\,-\,b^4}{\left(2a^2-b^2\right)S - 2a}} \label{eq:oblate_spheroid_rr_block}
     \end{align}
     \end{subequations}
     where 
     \begin{equation} \label{eq:oblate_spheroid_S_term}
         S = \frac{2}{\sqrt{b^2\,-\,a^2}}\,\arctan\left(\frac{\sqrt{b^2\,-\,a^2}}{a}\right)
     \end{equation}
     
    \vspace{0.5cm}
\item \textbf{Prolate Spheroid} $\left(a > b=c\right)$:\cite{Perrin1934,Koenig1975} \label{itm:prolate_spheroid}\\
The same analytical expressions in Eqs.~\eqref{eq:oblate_spheroid_tt_block} and~\eqref{eq:oblate_spheroid_rr_block}, but with
\begin{equation} \label{eq:prolate_spheroid_S_term}
         S = \frac{2}{\sqrt{a^2\,-\,b^2}}\,\ln\left(\frac{a+\sqrt{a^2\,-\,b^2}}{b}\right)
\end{equation}

     \vspace{0.5cm}
\item \textbf{Disk} $\left(a\to0 < b=c\right)$:\cite{Berg1993,Landau1987} \label{itm:disk}\\
Taking $a\to0$ for the oblate spheroid in item~\ref{itm:oblate_spheroid}, we have $\displaystyle{S=\frac{\pi}{b}}$ and
\begin{subequations}
    \label{eq:disk_blocks}
    \begin{equation}
    \xi^{\mathrm{tt}}_1 = \displaystyle{16\eta\,b} \qquad \xi^{\mathrm{tt}}_2 = \xi^{\mathrm{tt}}_3 = \displaystyle{\frac{32\eta}{3}\,b} \label{eq:disk_tt_block}
    \end{equation}
    \begin{equation}
    \xi^{\mathrm{rr}}_1 = \xi^{\mathrm{rr}}_2 = \xi^{\mathrm{rr}}_3 = \displaystyle{\frac{32\eta}{3}\,b^3} \label{eq:disk_rr_block}
    \end{equation}
     \end{subequations}
\end{enumerate}

\begin{figure}[h]
    \vspace{0.7cm}
    \centering
    \includegraphics[width=\linewidth]{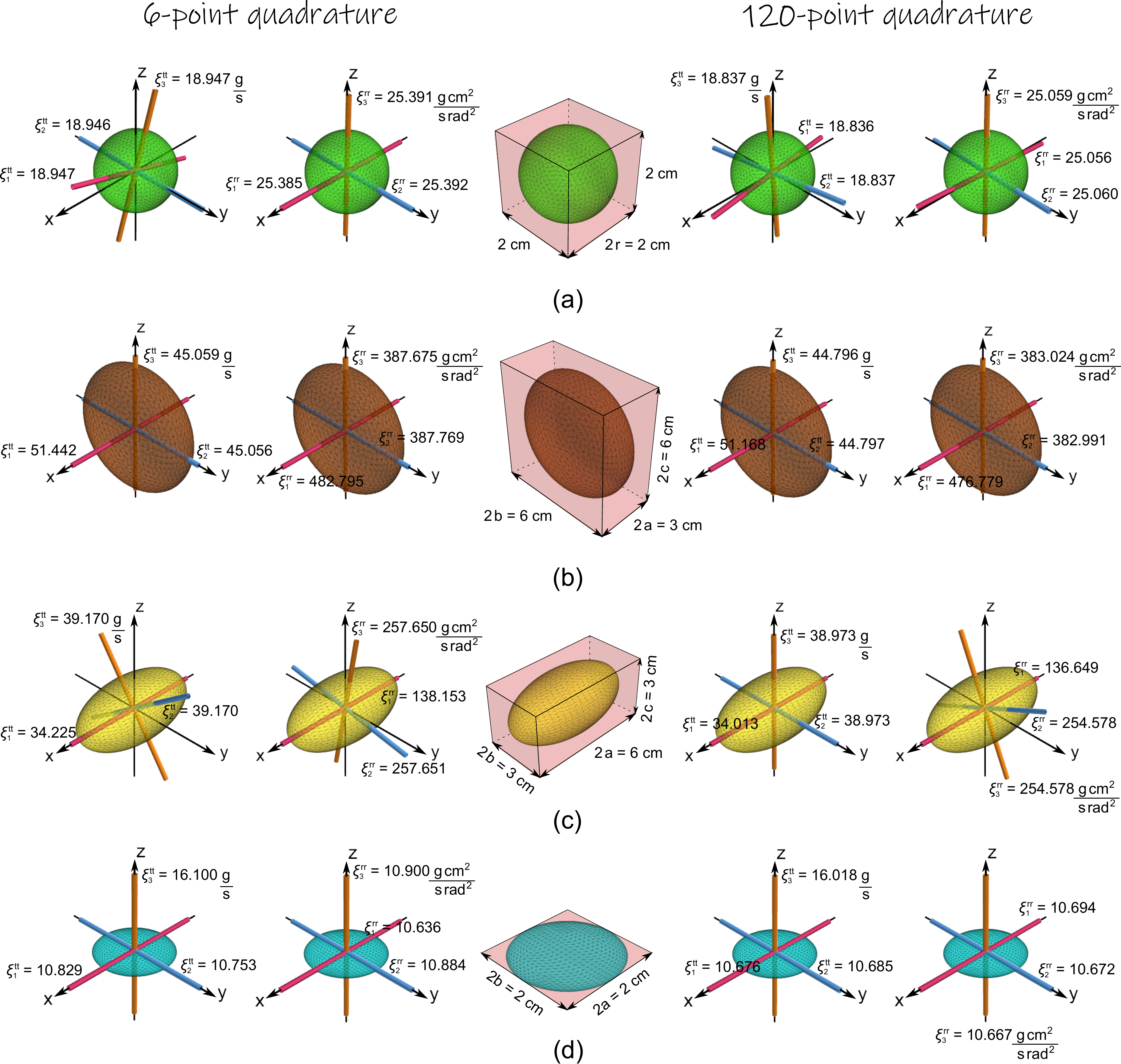}
    \caption{Moments and principal axes of the (tt) and (rr) blocks of the resistance tensor at the center of resistance using the boundary element method in Sec.~\ref{A-subsec:triangulated_surface_method}.  The moments of the (tt) block are in $\mathrm{g/s}$ and the moments of the (rr) block are in $\mathrm{(g~cm^2)/(s~rad^2)}$.  The two left columns employ a 6-point quadrature developed by Cowper\cite{Cowper1973}, and the two right columns employ a 120-point quadrature developed by Xiao and Gimbutas\cite{Xiao2010}. The panel in the middle displays the dimensions of the corresponding geometrical shapes: (a) Sphere, (b) Oblate Spheroid, (c) Prolate Spheroid and (d) Disk. For the degenerate moments, a linear combination of the associated axes will also form a basis.}
    \label{fig:analytical_geometries}
\end{figure}

\vspace{0.5cm}
To compare the analytical expressions in items~\ref{itm:sphere}--$\,$\ref{itm:disk} with the boundary element method developed in the main paper, we triangulated a sphere with radius 1~cm using the MSMS algorithm\cite{Sanner96} with a probe radius of 5.0~cm and a triangulation density of 100~$\mathrm{vertices/cm}^2$. Utilizing this sphere triangulation, we constructed the oblate and prolate spheroids and the disk by scaling the appropriate coordinates of the triangles' vertices. Figure~\ref{fig:analytical_geometries} shows the geometrical shapes and their dimensions associated with~Eq.~\eqref{eq:ellipsoid_equation}.

Figure~\ref{fig:analytical_geometries} also shows the moments and the principal axes of the (tt) and (rr) blocks. These were computed employing the boundary element method in Sec.~\ref{A-subsec:triangulated_surface_method} with the two integration quadratures utilized in this work and with $\eta=100$~mPa$\cdot$s. From Eq.~\eqref{A-eq:resistance_tensor}, the principal axes are associated with linear motion for the (tt)~block and rotational motion for the (rr)~block. The moments can be compared with the analytical values in Table~\ref{tab:analytical_moments_tt_and_rr_blocks}. The average percentage error for the 6-point quadrature is 0.86\% and for the 120-point quadrature, 0.15\%.    

\begin{table}[t]
    \caption{Analytical moments of the (tt) and (rr) blocks using the geometrical shapes in Fig.~\ref{fig:analytical_geometries}. We used the analytical expressions in the list of items~\ref{itm:sphere}--$\,$\ref{itm:disk} with $\eta=100$~mPa$\cdot$s. The blocks are at the center of resistance~(Sec.~\ref{A-subsec:center_of_pitch}) and follow the form in Eq.~\eqref{eq:geometries_analytical_form}.\label{tab:analytical_moments_tt_and_rr_blocks}}
    \centering
\begin{tabularx}{\textwidth} { 
   |c|
   >{\centering\arraybackslash}X 
  | >{\centering\arraybackslash}X 
  | >{\centering\arraybackslash}X 
  | >{\centering\arraybackslash}X 
  | >{\centering\arraybackslash}X 
  | >{\centering\arraybackslash}X 
  | >{\centering\arraybackslash}X 
  | >{\centering\arraybackslash}X 
  | >{\centering\arraybackslash}X |}
\hline 
\multirow{2}{*}{Object} & \multicolumn{3}{c|}{Dimensions~$\left(\mathrm{cm}\right)$} & \multicolumn{3}{c|}{tt block~${\left(\mathrm{g/s}\right)}$} & \multicolumn{3}{c|}{rr block~$\left(\frac{\mathrm{g~cm^2}}{\mathrm{s~rad^2}}\right)$}\\
\cline{2-10}
 & $a$ & $b$ & $c$ & $\xi^{\mathrm{tt}}_1$ & $\xi^{\mathrm{tt}}_2$ & $\xi^{\mathrm{tt}}_3$ & $\xi^{\mathrm{rr}}_1$ & $\xi^{\mathrm{rr}}_2$ & $\xi^{\mathrm{rr}}_3$\tabularnewline
\hline 
Sphere & \multicolumn{3}{c|}{1} & \multicolumn{3}{c|}{18.850} & \multicolumn{3}{c|}{25.133}\tabularnewline
\hline 
Oblate Spheroid & 1.5 & \multicolumn{2}{c|}{3} & 51.194 & \multicolumn{2}{c|}{44.827} & 478.415  & \multicolumn{2}{c|}{383.954 }\tabularnewline
Prolate Spheroid & 3 & \multicolumn{2}{c|}{1.5} & 34.041 & \multicolumn{2}{c|}{38.987 } & 136.849  & \multicolumn{2}{c|}{255.305 }\tabularnewline
\hline 
Disk\footnote{Because $c\to0<a=b$, the eigenvalues 1 and 3 are swapped when compared with the analytical expressions in item~\ref{itm:disk}.} & \multicolumn{2}{c|}{1} & 0 & \multicolumn{2}{c|}{10.667 } & 16.000 & \multicolumn{3}{c|}{10.667}\tabularnewline
\hline \hline
\end{tabularx}
\end{table}

\clearpage
\section{Helices \label{sec:helices}}

\begin{figure}[t]
    
    \includegraphics[width=0.7\linewidth]{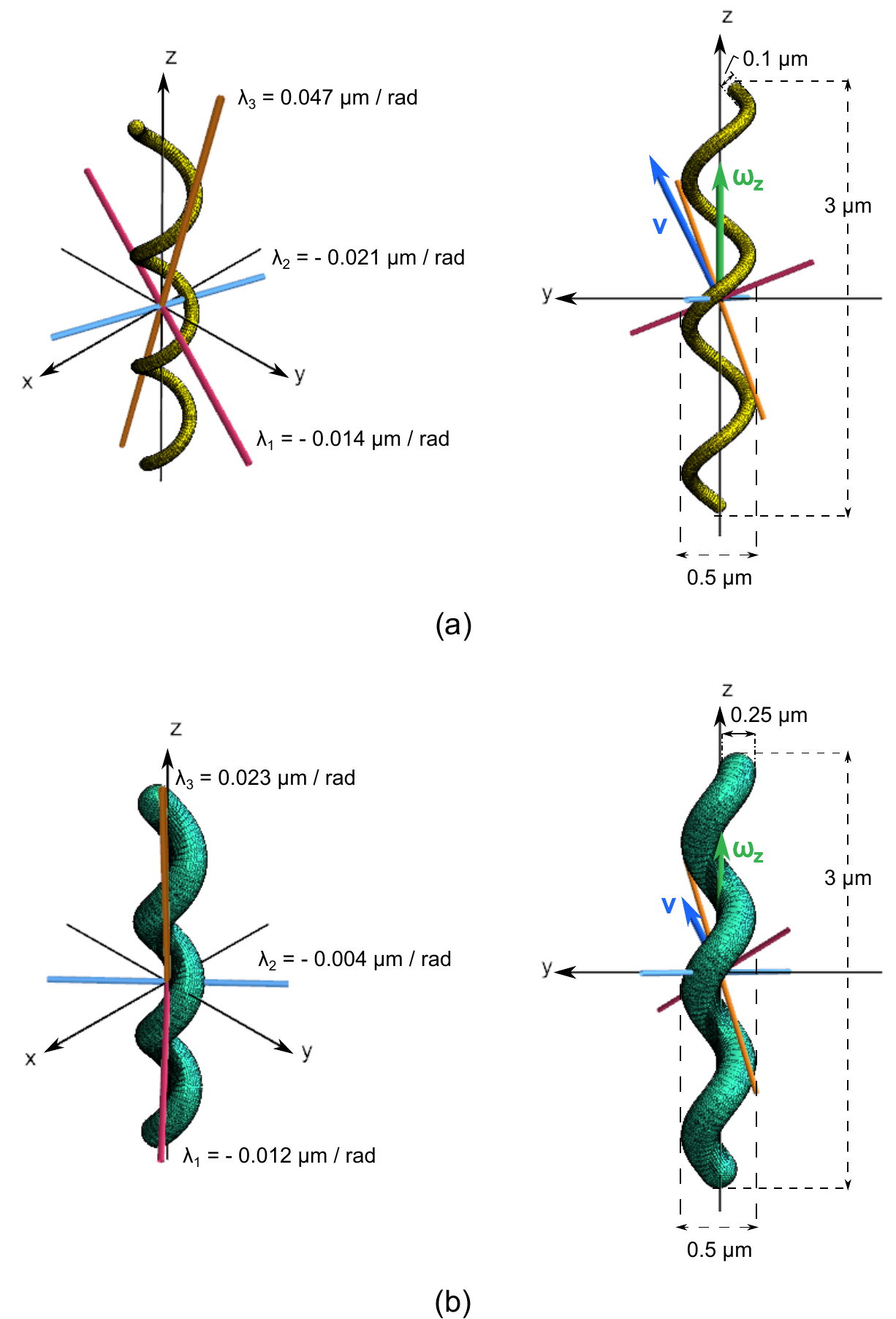}
    \caption{Pitch axes for two right-handed helices: (a) thickness = $0.1~\mu$m and (b) thickness = $0.25~\mu$m. The left panel displays the principal axes of pitch (pink, blue and orange lines) and the three associated moments of pitch for motion around those axes. The right panel displays the design parameters and the velocity vector $\mathbf{v}$ due to rotation around the $z$-axis with an angular velocity $\bm{\omega}_z$. \label{fig:helices_triangles}} \hypertarget{fig:helices_triangles}{}
\end{figure}

 To represent the helices in the Patil \textit{et al}.'s experiments\cite{Patil2021}, we have modeled helices with three flagellar widths: 0.1, 0.2 and 0.25~$\mu$m. Figure~\ref{fig:helices_triangles} shows the design parameters for the right-handed helices with thicknesses 0.1 (yellow) and 0.25~$\mu$m (turquoise) along with their pitch axes and associated moments of pitch. The main paper contains the helix with thickness 0.2~$\mu$m (red) (Fig.~\ref{A-fig:helix_triangles} in the main paper), and we built the other two helices with thicknesses 0.1 and 0.25~$\mu$m by following the same procedure outlined in the paper. The only difference is that the probe radius follows the helix's thickness and that the center of consecutive beads are 0.0245~$\mu$m apart for the helix with thickness 0.1~$\mu$m.     

Table~\ref{tab:helices_properties} displays the properties of the three right-handed helices. These helices have a projected pitch along the $z$-axis in the range 280--138~nm and can be compared to the reported experimental projected pitch of 250 nm by Patil \textit{et al}.\cite{Patil2021} Considering the conditions when translation-rotation coupling overtakes translational diffusion  (Eq.~\eqref{A-eq:condition_ratio_coupling_diffusion}), we can observe propulsion in a 100-second experiment when the helices have a minimum threshold frequency of 0.88 (yellow), 1.30 (red) and 1.67~Hz (turquoise helix). Patil \textit{et al}.\cite{Patil2021} also reported that their helices could overcome diffusion and propel when the rotation frequency was 2~Hz.  

\begin{table*}[h]

\caption{Properties of the helices used to model Patil \textit{et al}.'s experiments\cite{Patil2021}. The term $D_\mathrm{tt}$ is the translational diffusion coefficient, $|P|/{2\,\pi}$ (Eq.~\eqref{A-eq:pitch_coefficient}) is the scalar pitch coefficients and $\left(|\mathbf{v}|/|\bm{\omega}_z|\right) \, \times \, 2\pi$ is the pitch projected along the vector $\mathbf{v}$ after one complete revolution around the helices' long $z$-axes. The vector~$\mathbf{v}$ is computed using Eq.\eqref{A-eq:pitch_matrix_generalization} and $\bm{\omega}_z=\left(0,\,0,\,1\right)~\mathrm{rad~s}^{-1}$. The translational diffusion coefficients are based on dilute water solutions at 298.15 K with a viscosity of 0.89~mPa$\cdot$s.\cite{CRC_Handbook2022} Conditions when translation-rotation coupling overtakes translational diffusion are also provided (Eq.~\eqref{A-eq:condition_ratio_coupling_diffusion}). \label{tab:helices_properties} } 
\begin{tabular*}{\linewidth}{@{\extracolsep{\fill}}cccc}
\hline 
\multirow{2}{*}{Property} & \multicolumn{3}{c}{Helix} \\
\cline{2-4} \cline{3-4} \cline{4-4} 
 & yellow (Fig.~\ref{fig:helices_triangles}\protect\hyperref[fig:helices_triangles]{(a)}) & red (Fig.~\ref{A-fig:helix_triangles}) & turquoise (Fig.~\ref{fig:helices_triangles}\protect\hyperref[fig:helices_triangles]{(b)}) \\
\hline 
thickness ($\mu$m) & 0.1 & 0.2 & 0.25 \\
$\left(|\mathbf{v}|/|\bm{\omega}_z|\right) \, \times \, 2\pi$ (nm) & 280 & 180 & 138 \\
$D_\mathrm{tt}$ $\left(10^5~\mathrm{nm}^{2}~\mathrm{s}^{-1}\right)$ & 4.81 & 4.39 & 4.25 \\
$|P|/{2\,\pi}$ $\left(\mathrm{nm}~\mathrm{rad}^{-1}\right)$ & 30.8 & 19.9 & 15.2 \\
$|\bm{\omega}|^{2}\times t$ $\left(10^{3}~\mathrm{rad}^{2}~\mathrm{s}^{-1}\right)$ & > 3.04 & > 6.65 & > 11.0 \\
\hline \hline
\end{tabular*}
\end{table*}

\clearpage
\section{Non-Swimmers and Achiral Swimmers \label{sec:non_swimmers_achiral_swimmers}}

\begin{figure}[h]
    \centering
    \includegraphics[width=\linewidth]{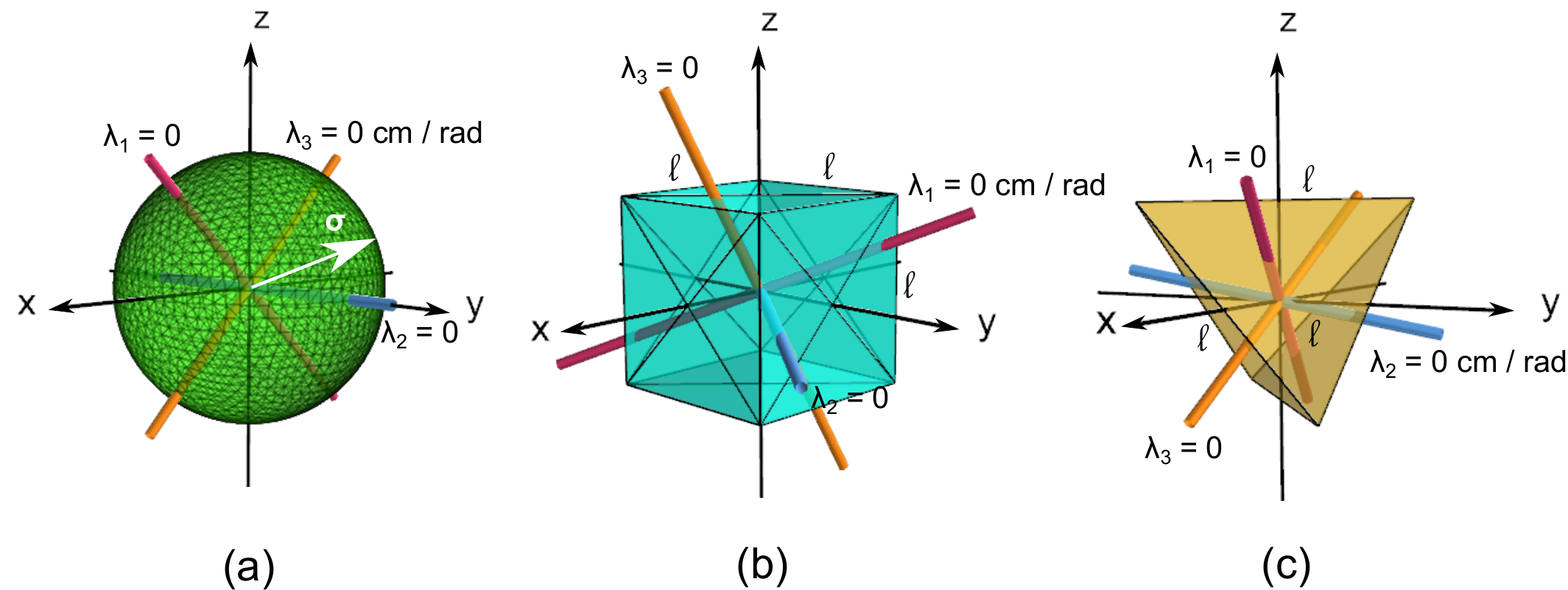}
    \caption{The principal axes of pitch and the associated moments of pitch for: (a) a sphere with radius $\sigma=2.2$~cm, (b) a cube with edge length {\textcursive{l}}~$\,=3~\mathrm{cm}$ and (c) a tetrahedron with {\textcursive{l}}~$\,=3\sqrt{2}~\mathrm{cm}$.  The moments of pitch are in $\mathrm{cm/rad}$.  Achiral objects with high degrees of symmetry (spheres, cubes, tetrahedra, and ellipsoids) have pitch matrices which are the null matrix, so all moments of pitch are zero. Because the moments of pitch are degenerate, the principal axes of pitch for these objects can be taken to be any three perpendicular axes, and there should be no coupling between rotational and translational motion in a fluid.}
    \label{fig:sphere_cube_tetrahedron}
\end{figure}

\begin{figure}[h]
    \centering
    \includegraphics[width=\linewidth]{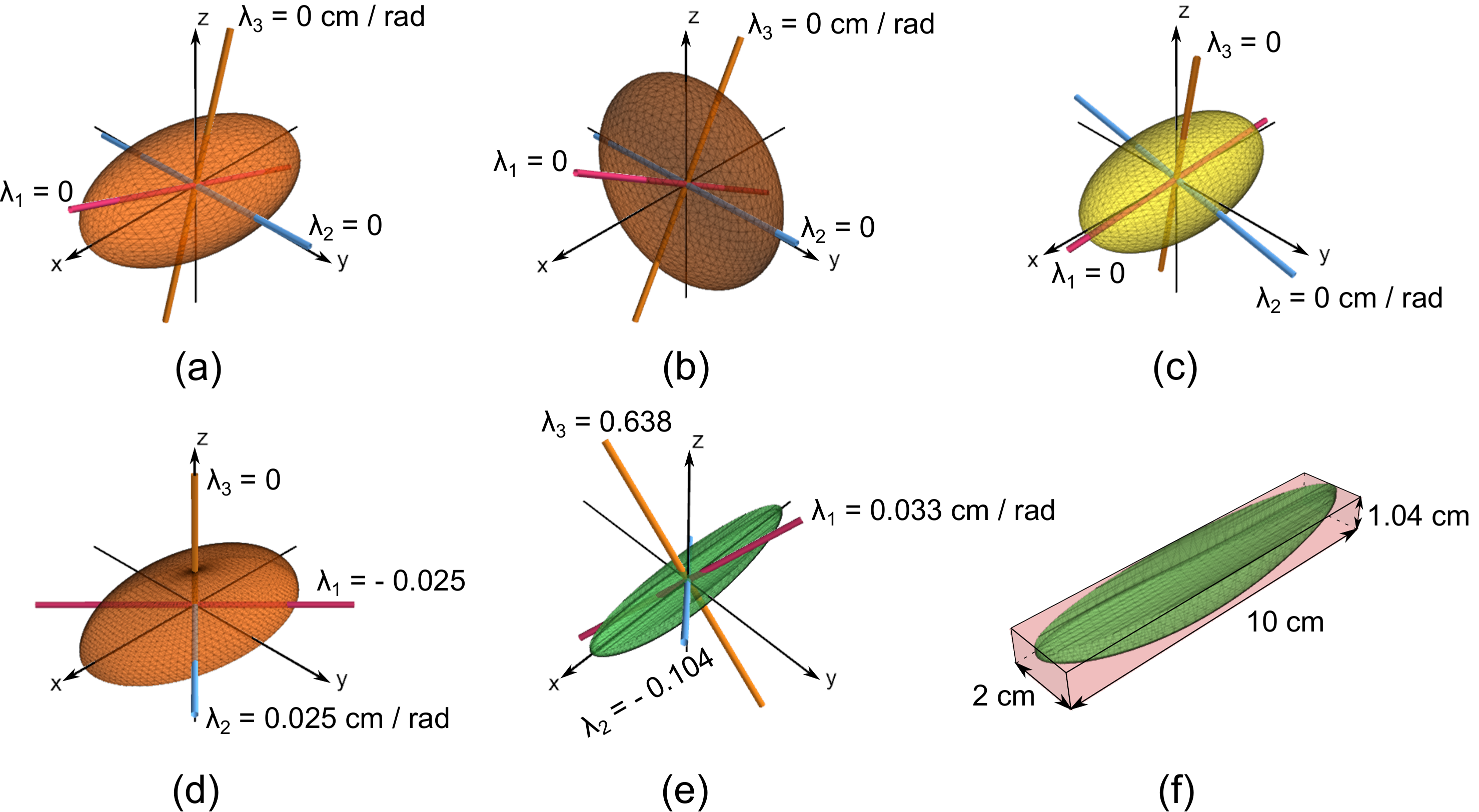}
    \caption{Like other high symmetry objects, ellipsoids do not exhibit rotation-translation coupling. The examples in the top row are triaxial (a), oblate (b), and prolate (c) ellipsoids. Using Eq.~\eqref{eq:ellipsoid_equation}, the semi-axes $\left(a,\,b,\,c\right)$  for these ellipsoids are: $\left(5.0,\,3.0,\,2.0\right)$, $\left(1.5,\,3.0,\,3.0\right)$, and $\left(3.0,\,1.5,\,1.5\right)$, respectively (all in cm). The bottom row shows partial ellipsoids: (d) is the upper half of the triaxial ellipsoid (a semi-ellipsoid) which is an achiral swimmer, while (e) and (f) show the asymmetric ``rattleback'' which prefers one rotational direction when it is spun on a planar surface. The axes of pitch shown here do not govern the rattleback's behavior on a solid surface, but they indicate that it should have rotational-translational coupling in a fluid.  The moments of pitch are in $\mathrm{cm/rad}$.  \label{fig:ellipsoids_rattleback}}
    \hypertarget{fig:ellipsoids_rattleback}{}
\end{figure}

Non-swimming achiral objects have null pitch matrices and moments of pitch that are all zero. In contrast, achiral swimmers have one moment of pitch that is zero and two other moments that have the same magnitude with a flipped sign. Figure~\ref{fig:sphere_cube_tetrahedron} displays examples of non-swimming achiral objects. To triangulate the surface of the sphere, we used the MSMS algorithm\cite{Sanner96} with a probe radius of 1.4~cm and a triangulation density of 25~$\mathrm{vertices/cm}^2$. Each cube face was divided into 4 equal triangles and the tetrahedron is already formed by 4 triangles. Some other examples of non-swimming achiral objects are the ellipsoids in the upper panels of Fig.~\ref{fig:ellipsoids_rattleback}. These non-swimming objects all have a high degree of internal symmetry.

The ellipsoids in the upper panels of Fig.~\ref{fig:ellipsoids_rattleback} were constructed by scaling the vertices' coordinates of a triangulated sphere with radius 1 cm. To triangulate the surface of the unit sphere, we used the MSMS algorithm\cite{Sanner96} with a probe radius of 5.0 cm and a triangulation density of 100~$\mathrm{vertices/cm}^2$. Using Eq.~\eqref{eq:ellipsoid_equation}, the scaling factors $\left(a,\,b,\,c\right)$ are: $\left(5.0,\,3.0,\,2.0\right)$ for the triaxial ellipsoid, $\left(1.5,\,3.0,\,3.0\right)$ for the oblate spheroid and $\left(3.0,\,1.5,\,1.5\right)$ for the prolate spheroid. For all objects in Fig.~\ref{fig:ellipsoids_rattleback}, we employed the boundary element method described in Sec.~\ref{A-subsec:triangulated_surface_method} with a numerical integration that exactly integrates polynomials of degree 25.\cite{Xiao2010} This numerical integration is available in Quadpy\cite{Quadpy_github} as the ``Xiao-Gimbutas 25'' quadrature.\cite{Xiao2010}   

The semi-ellipsoid in the lower left of Fig.~\ref{fig:ellipsoids_rattleback} is an example of an achiral swimmer. It is the upper half of the triaxial ellipsoid in Fig.~\ref{fig:ellipsoids_rattleback}\protect\hyperlink{fig:ellipsoids_rattleback}{(a)} (sliced along the $xy$-plane), and was modeled and triangulated using OpenSCAD.\cite{OpenSCAD} With minor modifications from a semi-ellipsoid, a right-handed rattleback (which prefers to rotate counterclockwise on a planar surface) was constructed following the method of Ref.~\onlinecite{Elliott2010} and triangulated using OpenSCAD.\cite{OpenSCAD} The rattleback is chiral and, from its pitch axis and moments of pitch, it should not have a preferred direction of rotation when immersed in a fluid. This is distinct from its behavior on a solid planar surface. On a solid planar surface, however, we note that its largest moment of pitch indicates that a counterclockwise rotation would act in the opposite direction of the surface (preferred), while a clockwise rotation would act against the surface (by providing more friction). For the semi-ellipsoid, one moment of pitch is zero and it should not exhibit the same behavior as the rattleback when spun.

\clearpage
\section{Isotropic Helicoids \label{sec:isotropic_helicoids_sm}}

\begin{figure}[b]
    \includegraphics[width=0.6\linewidth]{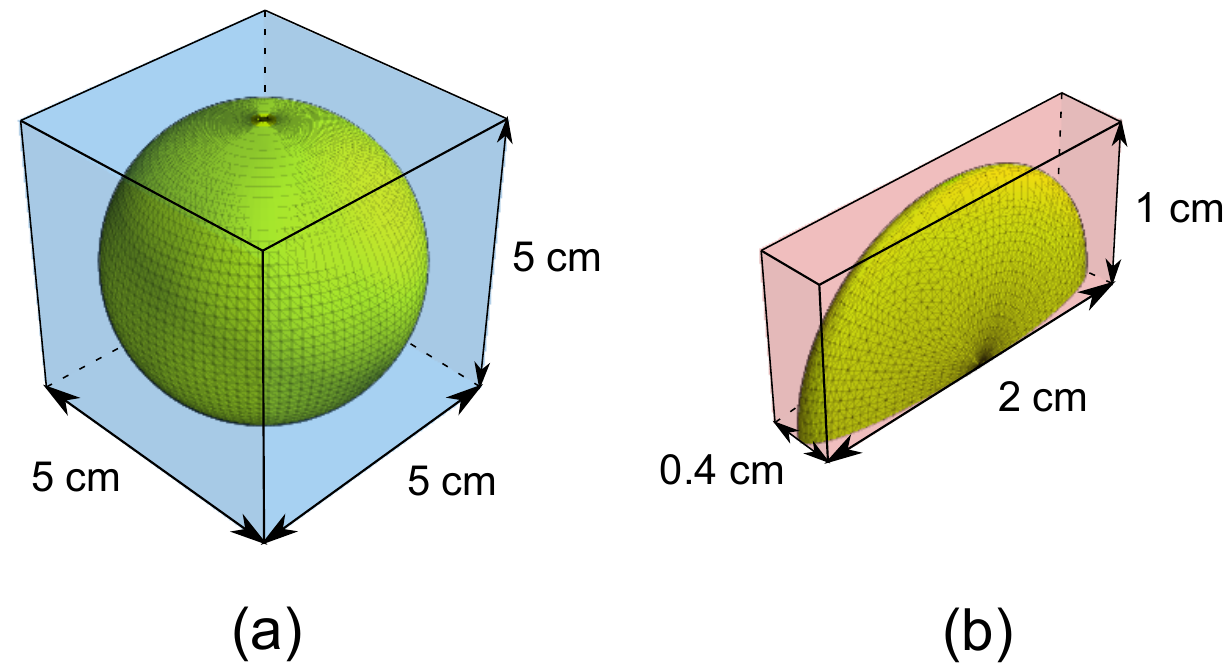}
    
    \caption{Generalization of Lord Kelvin's Isotropic Helicoid. (a) Vanes are placed midway between the intersection points of three circles around a sphere. (b) The vanes are tilted at an angle $\theta$ relative to the circles, and this angle defines the handedness of the isotropic helicoid. The dimensions of (c) the sphere and (d) the semi-oblate vanes used in our examples. \label{fig:isotropic_helicoids_construction_SI}}
    \hypertarget{fig:isotropic_helicoids_construction_SI}{}
\end{figure}

\begin{figure}[t]
    \includegraphics[width=0.92\linewidth]{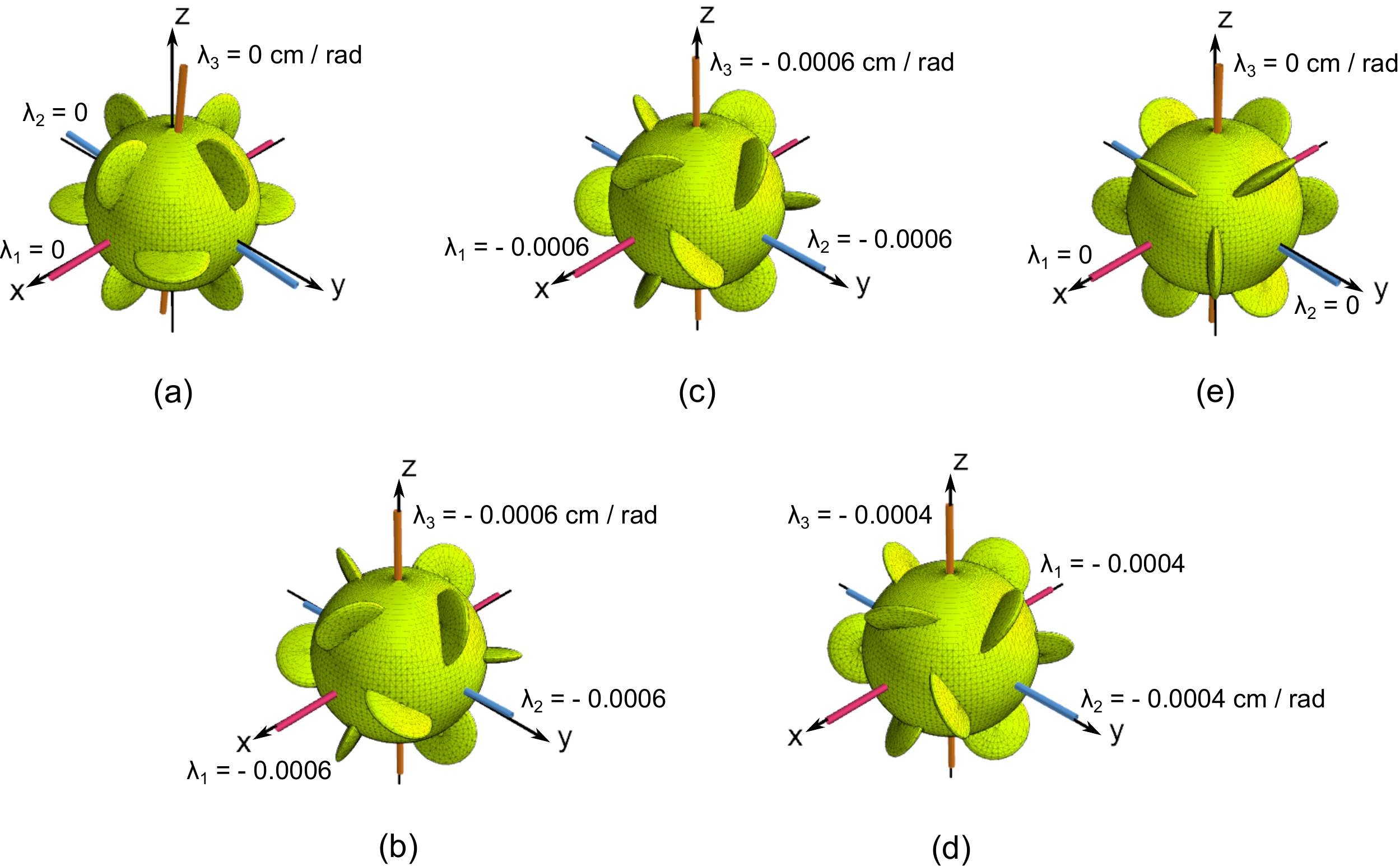}
    
    \caption{ Moments of pitch and pitch axes for spherically isotropic bodies (panels (a) and (e)) and left-handed \textit{isotropic helicoids} (panels (b), (c) and (d)). The moments of pitch $\lambda_i$ are in $\mathrm{cm/rad}$. The vane angles are: (a) $0^\circ$, (b) $-30^\circ$, (c) $-45^\circ$, (d) $-60^\circ$ and (e) $-90^\circ$. Our numerical calculations of the pitch axes and moments of pitch indicate that only the isotropic helicoids have non-zero moments of pitch. Each of these isotropic objects is composed of a sphere with 12 semi-oblate vanes whose dimensions are shown in Fig.~\ref{fig:isotropic_helicoids_construction_SI}.}
    \label{fig:kelvin_SI}
    \hypertarget{fig:kelvin_SI}{}
\end{figure}

 We constructed three other left-handed isotropic helicoids $(\theta = -30^\circ, -45^\circ, -60^\circ)$ and two other spherically isotropic bodies $(\theta = 0^\circ, -90^\circ)$ using the same procedure in the main paper. The dimensions of the sphere and the semi-oblate vanes are in Fig.~\ref{fig:isotropic_helicoids_construction_SI}. When compared to the set of isotropic helicoids and bodies from the main paper, this new set is made with the same sphere but with slightly smaller semi-oblate vanes. Figure~\ref{fig:kelvin_SI} shows the moments of pitch and pitch axes for the new set of isotropic helicoids and bodies. The moments of pitch are lower when the semi-oblate vanes are smaller. This suggests that the vanes significantly contribute to the rotation-translation coupling of the isotropic helicoids. 

Table~\ref{tab:isotropic_helicoids_blocks_resistance_tensors_SI} shows the scalar parameters (Eq.~\ref{A-eq:scalar_helicoid}) for the translational (tt), translation-rotation (tr), and rotational (rr) blocks of the resistance tensor for the new set of isotropic helicoids and bodies. It also shows the terminal velocities, $\mathrm{v}$, and terminal angular velocities, $\omega$, for these bodies as they fall through a viscous fluid, using the same experimental conditions in the main paper. When compared to the bodies in the main paper, the scalar parameters and terminal angular velocities are lower, but the terminal velocities are practically the same.

\begin{table}[b]

\caption{Scalar parameters (Eq.~\ref{A-eq:scalar_helicoid}) for the (tt), (tr), and (rr) blocks of the resistance tensor for isotropic helicoids with different vane angles $(\theta)$. The moments of pitch for these objects can be obtained from $-\,\xi^\mathrm{tr} / \xi^\mathrm{tt}\,$. Also shown are the terminal velocities, $\mathrm{v}$, and terminal angular velocities, $\omega$, for these bodies as they fall through a viscous fluid, using the same experimental conditions in the main paper.
\label{tab:isotropic_helicoids_blocks_resistance_tensors_SI}}

\begin{tabular*}{\linewidth}{@{\extracolsep{\fill}}cccccc}
\hline 
\multirow{2}{*}{Angle}  & $\xi^{\mathrm{tt}}$  & $\xi^{\mathrm{tr}}$  & $\xi^{\mathrm{rr}}$ & $\mathrm{v}$ & $\omega$  \\ 
 & $\left(\mathrm{kg/s}\right)$ &  $\left(\mathrm{kg~m/(s~rad)} \right)$ & $\left(\mathrm{kg~m^2/(s~rad^2)}\right)$  & ($\mathrm{cm/s}$) & ($\mathrm{rad/s}$)  \\ \hline
$\phantom{-}0^\circ$ & 0.264 & 0 & $2.99\times 10^{-4}$ & 47.1 & 0 \\
$-30^\circ$ & 0.264  & $1.55 \times 10^{-6}$& $2.99\times 10^{-4}$  & 47.1 & $-2.44\times 10^{-3}$ \\
$-45^\circ$ & 0.264 & $1.57 \times 10^{-6}$& $2.99\times 10^{-4}$  & 47.1 & $-2.48\times 10^{-3}$ \\
$-60^\circ$ & 0.264  & $1.11 \times 10^{-6}$& $2.99\times 10^{-4}$  & 47.1 & $-1.75\times 10^{-3}$ \\
$-90^\circ$ & 0.264  & 0 & $2.99\times 10^{-4}$ & 47.1 &  0 \\
\hline \hline
\end{tabular*}
\end{table}

\begin{figure}[h]
    \vspace{0.7cm}
    \centering
    \includegraphics[width=\linewidth]{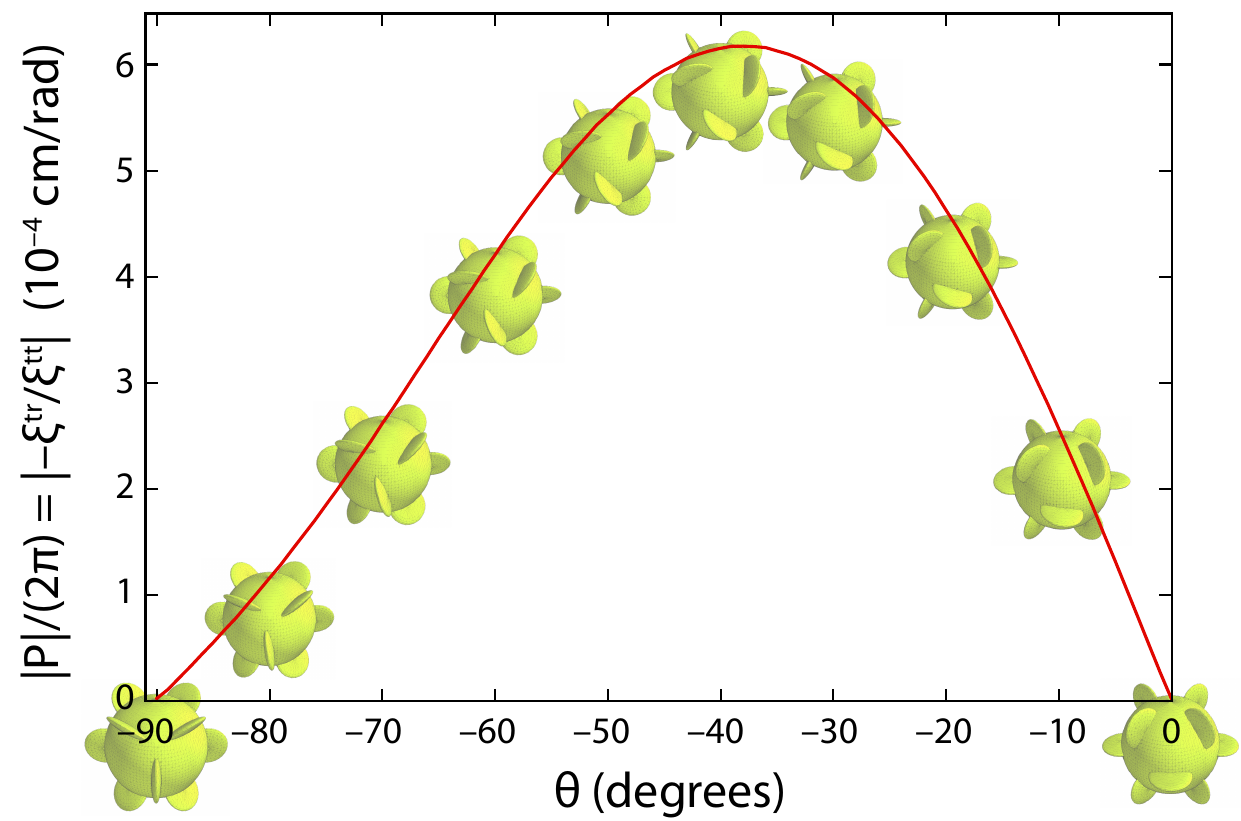}
    \caption{Pitch Coefficient, $\left| P\right|/\left(2 \pi\right)$, for Lord Kelvin's isotropic helicoids composed of a sphere with 12 semi-oblate vanes whose dimensions are shown in Fig.~\ref{fig:isotropic_helicoids_construction_SI}. The vane angle $\theta$ was changed in $1^{\circ}$ increments to generate this plot. Representative isotropic bodies are shown in $10^{\circ}$ increments. For $\theta=0^{\circ}$ and $\theta=-90^{\circ}$, we recover spherically isotropic bodies which do not couple rotational and translational motion.}
    \label{fig:isotropic_helicoid_varying_angle_plot}
\end{figure}

Figure~\ref{fig:isotropic_helicoid_varying_angle_plot} displays the pitch coefficients (Eq.~\eqref{A-eq:pitch_coefficient}) for Lord Kelvin's isotropic helicoids after varying the angle $\theta$ from $0^{\circ}$ to $-90^{\circ}$. These isotropic helicoids are constructed with the sphere and the semi-oblate vanes in Fig.~\ref{fig:isotropic_helicoids_construction_SI}. 
The pitch coefficient has a maximum value when $\theta=-38^{\circ}$.   

\clearpage
\section{The Relationship between Pitch and Moment of Inertia for Spheres \label{sec:relationship_pitch_moment_inertia}}

Consider a sphere with radius $R$. At the center of resistance (CR), which is the center of the sphere\cite{Torre:1983lr,Duraes2021}, we know that the coupling blocks of the resistance tensor are null matrices (see Sec.~\ref{sec:analytical_expressions}). From the definition of pitch matrix in Eq.~\eqref{A-eq:pitch_matrix_resistance_tensor}, the pitch matrix at (CR) is also a $3 \times 3$ null matrix. From Sec.~\ref{sec:proof_uniqueness_center_of_pitch}, note that the center of pitch is the same as (CR) for the sphere.

One interesting observation is that, using the pitch matrix at (CR), we can find the pitch matrix at an arbitrary point $M$ from Eq.~\eqref{A-eq:pitch_matrix_transform}:
\begin{align}
\begin{split}
\label{eq:sphere_pitch_matrix_arbitrary_point}
\frac{\mathsf{P}_{M}}{2\pi} & = \frac{\mathsf{P}_{\mathrm{CR}}}{2\pi} - \mathsf{U}_{M} \\
\frac{\mathsf{P}_{M}}{2\pi} & = -\,\mathsf{U}_{M} 
\end{split}
\end{align}
since $\displaystyle{\frac{\mathsf{P}_{\mathrm{CR}}}{2\pi}=\mathbf{0}}$.

\vspace{0.05in}
We take the center of the sphere to be the origin $\left(0,\,0,\,0\right)$ and write the point $M$ in spherical coordinates,
\begin{equation} \label{eq:point_M_spherical_coordinates}
    M = \left(r\sin{\theta}\cos{\phi},\,r\sin{\theta}\sin{\phi},\,r\cos{\theta}\right)\,,
\end{equation}
where $r\geq0$, $0 \leq \theta \leq \pi$ and $0 \leq \phi < 2\pi$.

This way, the pitch matrix for the sphere at a point $M$ is:
\begin{align} \label{eq:pitch_matrix_point_M_spherical_coordinates}
    \frac{\mathsf{P}_{M}}{2\pi} = -\,\left[ \begin{array}{ccc}
  \phantom{-\,}0 & -\,r\cos{\theta} & \phantom{-\,}r\sin{\theta}\sin{\phi} \\
\phantom{-\,}r\cos{\theta} &  \phantom{-\,}0   & -\,r\sin{\theta}\cos{\phi} \\
-\,r\sin{\theta}\sin{\phi} &  \phantom{-\,}r\sin{\theta}\cos{\phi} & \phantom{-\,}0
\end{array} \right] 
\end{align}
where the skew-symmetric matrix $\displaystyle{\mathsf{U}_{M}}$ in Eq.~\eqref{eq:sphere_pitch_matrix_arbitrary_point} is defined in Eq.~\eqref{A-eq:skew_symmetric_matrix} in the main paper.

Because the pitch matrix in Eq.~\eqref{eq:pitch_matrix_point_M_spherical_coordinates} has three distinct eigenvalues, it is diagonalizable.\cite{Riley_Hobson_Bence2006} The three eigenvalues are $\lambda_1 = 0$ and the complex conjugates $\lambda_2 = ri$ and $\lambda_3=\left(\lambda_2\right)^{*}=-\,ri$, which are all independent of the angles $\theta$ and $\phi$. Note that if the point $M$ is at the center of the sphere (\textit{i.e.}, $r=0$), we recover the result from Fig.~\ref{fig:sphere_cube_tetrahedron}. 

From Sec.~\ref{sec:pitch_coefficient_generalization}, we can compute the pitch coefficient, 
\begin{equation} \label{eq:pitch_coefficient_sphere_point_M}
    \frac{|P|}{2 \pi} = \sqrt{\frac{2}{3}\;r^2}
\end{equation}

If the point $M$ is at the surface of the sphere (\textit{i.e.}, $r=R$) and the sphere is rotating with angular velocity $\omega$, we can equate linear kinetic energy to rotational kinetic energy and find the moment of inertia $\left(I\right)$:
\begin{align} \label{eq:sphere_pitch_and_moment_of_inertia}
    \begin{split}
    \frac{1}{2}\,m\,\mathrm{v}^2 & = \frac{1}{2}\,I\,\omega^2 \\
    m\,\left(\frac{|P|}{2 \pi}\;\omega\right)^2 & = I\,\omega^2 \\
    \implies I & = m\,\left(\frac{|P|}{2 \pi}\right)^2 = \frac{2}{3}\;m\,R^2
    \end{split}
\end{align}
where $m$ is the mass and the velocity $\left(\mathrm{v}\right)$ comes from the pitch coefficient and the angular velocity. This is the moment of inertia of a thin uniform spherical shell about any axis through its center.\cite{Morin2008} Note that if $R=0$, we have a point mass rotating around its center, and then $I=0$.

The result in Eq.~\eqref{eq:sphere_pitch_and_moment_of_inertia} agrees with the observation that only the region in contact with the fluid (the sphere's surface) contributes to hydrodynamic interactions.\cite{Garcia1999}  This puts the boundary element method developed in the main paper on firmer footing for predicting hydrodynamic interactions with the surrounding fluid.

\clearpage
\section{Spheres in a Non-Newtonian Fluid}

In two different non-Newtonian fluids, Rogowski~\textit{et al.}\cite{Rogowski2021} reported translation-rotation coupling for magnetic microspheres. They measured propulsion velocities whose origins were the rotational motion of the magnetic spheres. From Sec.~\ref{sec:relationship_pitch_moment_inertia} above, we know that this is possible when the center of pitch is not coincident with the center of the sphere, \textit{i.e.}, when the pitch matrix is no longer symmetric.   

Utilizing the experimental data available in the Supporting Information of Ref.~\onlinecite{Rogowski2021}, we can estimate the `symmetry-breaking point' where the pitch matrix is no longer symmetric. For a sphere of radius $R$ centered at the origin, we can use the pitch coefficient in Eq.~\eqref{eq:pitch_coefficient_sphere_point_M} to compute the propulsion velocity $\left(\mathrm{v}\right)$ that arises solely from angular velocity $\left(\omega\right)$: 
\begin{align} \label{eq:velocity_sphere_point_M}
\begin{split}
    \mathrm{v} & = \frac{|P|}{2 \pi}\; \omega \\
      & = \left(\sqrt{\frac{2}{3}\;r^2}\right) 2\pi f \\
      & = \left(2\pi\sqrt{\frac{2}{3}}\right) \alpha R\,f
\end{split}
\end{align}
where $f$ is the rotation frequency, $\alpha R$ is the location of the symmetry-breaking point, and $\alpha$ is a non-negative dimensionless parameter. From Sec.~\ref{sec:relationship_pitch_moment_inertia}, we note that the symmetry-breaking point is actually a set of points on the surface of a sphere with radius $\alpha R$, since it is independent of the spherical-coordinate angles $\theta$ and $\phi$.

\begin{figure}[h]
    \vspace{0.7cm}
    \centering
    \includegraphics[width=\linewidth]{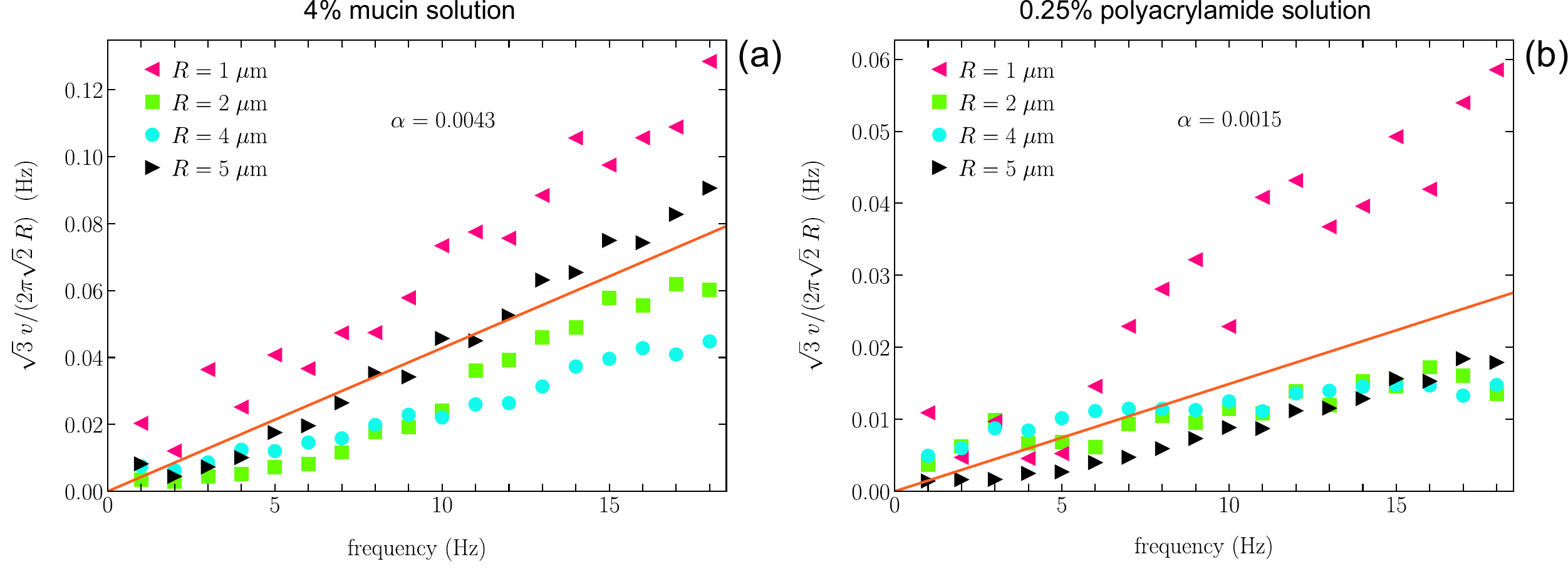}
    \caption{Linear fit (orange curve) to estimate the parameter $\alpha$ in Eq.~\eqref{eq:velocity_sphere_point_M} for spheres with radii~$R$ in (a)~4\% mucin solution and (b)~0.25\% polyacrylamide solution. The experimental data comes from the overall velocities associated with Supplementary Fig.~16 of Ref.~\onlinecite{Rogowski2021}. For the  4\% mucin solution, the standard deviation is 0.0002 and the coefficient of determination $\left(\mathrm{R}^2\right)$ is 0.861. For the  0.25\% polyacrylamide solution, the standard deviation is 0.0001 and the coefficient of determination $\left(\mathrm{R}^2\right)$ is 0.713.  The values of $\alpha$ indicate that the point at which the spheres break the pitch matrix symmetry in these non-Newtonian fluids actually lie quite close to the center of pitch. \label{fig:parameter_alpha_4_mucin_and_0_25_polyacrylamide}}
\end{figure}

Figure~\ref{fig:parameter_alpha_4_mucin_and_0_25_polyacrylamide} displays the estimated parameter $\alpha$ from Eq.~\eqref{eq:velocity_sphere_point_M} for microspheres in two non-Newtonian fluids: 4\% mucin and 0.25\% polyacrylamide solutions. These two fluids require different $\alpha$ values, but both values indicate that the symmetry-breaking points lie closer to the center of the spheres than to the spheres' surfaces. For the 4\% mucin solution, the microspheres with radii equal to 2, 4 and 5~$\mu\mathrm{m}$ have a common parameter~$\alpha$ for low and mid-range rotation frequencies. For the 0.25\% polyacrylamide solution, the same microspheres have a common parameter~$\alpha$ throughout the rotation frequencies. Although the experimental data for the sphere with $R=1\,\mu\mathrm{m}$ shows that it has a distinct $\alpha$~value when compared to the others, it shares a common trend in the range 1--$\,$6 Hz for the 0.25\% polyacrylamide solution.  Overall, the experimental data suggests that spherical particles can have a general parameter $\alpha$ that characterizes their symmetry-breaking points at low and mid-range rotation frequencies. 

We also note that if the viscosity $\eta$ is treated as a function, rather than a constant, in the Stokes expressions for spheres (Eq.~\eqref{eq:sphere_blocks}), we obtain the same pitch matrix as in Eq.~\eqref{eq:pitch_matrix_point_M_spherical_coordinates}.

\clearpage
\bibliography{main}